\documentclass{aa}  
\usepackage{graphicx}
\usepackage{txfonts}
\usepackage{mathpazo}
\usepackage[T1]{fontenc}
\usepackage{xcolor}
\usepackage{appendix}
\usepackage{multicol}
\usepackage[flushleft]{threeparttable}
\usepackage{array,booktabs}
\usepackage{amsmath}
\usepackage{textcomp}
\usepackage{hyperref}
\hypersetup{
    colorlinks=true,
    linkcolor=blue,
    filecolor=red,
    citecolor=blue,
    urlcolor=blue,
}

\begin{document} 

   \title{Linking ice and gas in the $\lambda$ Orionis Barnard 35A cloud}
   
     \author{G. Perotti
          \inst{1}\fnmsep,
          J. K. J{\o}rgensen\inst{1},
          H. J. Fraser\inst{2},
          A.N. Suutarinen\inst{2},
          L. E. Kristensen\inst{1},
          W. R. M. Rocha\inst{1},\\
          P. Bjerkeli\inst{3},
          \and
          K. M. Pontoppidan\inst{4}
        }

    \institute{Niels Bohr Institute \& Centre for Star and Planet Formation, University of Copenhagen, {\O}ster Voldgade 5$-$7, 1350 Copenhagen K., Denmark\\
              \email{giulia.perotti@nbi.ku.dk}
        \and
             School of Physical Sciences, The Open University, Walton Hall, Milton Keynes, MK7 6AA, United Kingdom
        \and
             Department of Space, Earth, and Environment, Chalmers University of Technology, Onsala Space Observatory, 439 92 Onsala, Sweden   
        \and
             Space Telescope Science Institute, 3700 San Martin Drive, Baltimore, MD 21218, USA
       }

   \date{Received ; accepted }

 
  \abstract
   {Dust grains play an important role in the synthesis of molecules in the interstellar medium, from the simplest species such as H$_2$ to complex organic molecules. How some of these solid-state molecules are converted into gas-phase species is still a matter of debate.}
   {Our aim is to directly compare ice and gas abundances of  methanol (CH$_3$OH) and carbon monoxide (CO), obtained from near-infrared (2.5$-$5~$\mu$m) and millimeter (1.3~mm) observations, and to investigate the relationship between ice, dust and gas in low-mass protostellar envelopes.}
   {We present Submillimeter Array (SMA) and Atacama Pathfinder EXperiment (APEX) observations of gas-phase CH$_3$OH ($J_K$~=~5$_K-$4$_K$), $^{13}$CO and C$^{18}$O ($J$~=~2$-$1) towards the multiple protostellar system IRAS~05417+0907 located in the B35A cloud, $\lambda$ Orionis region. We use archival IRAM~30~m data and AKARI H$_2$O, CO and CH$_3$OH ice observations towards the same target to compare ice and gas abundances and directly calculate CH$_3$OH and CO gas-to-ice ratios.}
    {The CO isotopologues emissions are extended, whereas 
    the CH$_3$OH emission is compact and traces the giant molecular outflow emanating from IRAS~05417+0907. A discrepancy between submillimeter dust emission and H$_2$O ice column density is found for B35A$-$4 and B35A$-$5, similar to what has previously been reported. B35A$-$2 and B35A$-$3 are located where the submillimeter dust emission peaks and show H$_2$O column densities lower than for B35A$-$4.}
   {The difference between the submillimeter continuum emission and the infrared H$_2$O ice observations suggests that the distributions of dust and H$_2$O ice differ around the young stellar objects in this dense cloud. The reason for this may be that the four sources are located in different environments resolved by the interferometric observations: B35A$-$2, B35A$-$3 and in particular B35A$-$5 are situated in a shocked region plausibly affected by sputtering and heating impacting the submillimeter dust emission pattern, while B35A$-$4 is situated in a more quiescent part of the cloud. Gas and ice maps are essential to connect small-scale variations in the ice composition with large-scale astrophysical phenomena probed by gas observations.}
    \titlerunning{Linking ice and gas in the the $\lambda$ Orionis Barnard 35A cloud}
    \authorrunning{G. Perotti et al.}

   \maketitle


\section{Introduction}
\label{Introduction}

The interaction between dust, ice and gas is ubiquitous in star-forming regions and it is essential for the synthesis of interstellar molecules, the main ingredients for the origin of life on Earth. In the last decades the number of detected molecules in the interstellar medium (ISM) increased considerably \citep{McGuire2018} and with it, the awareness of an interplay between solid and gaseous molecules in star-forming regions \citep{Herbst2009,Boogert2015,Jorgensen2020,Oberg2020}. Some of the questions remained to be answered are what is the link between the distribution of solids (dust, ices) and gaseous molecules in molecular clouds, and what does this tell us about the solid-gas intertwined chemistries (e.g., thermal and non-thermal desorption mechanisms releasing solid-state molecules into the gas-phase and vice versa).

Desorption mechanisms are of utmost importance to understand some critical aspects of star- and planet-formation \citep{vanDishoeck1998}. Besides enhancing the chemical complexity in the gas-phase \citep{Cazaux2003,Jorgensen2016,Bergner2017,Calcutt2018,Manigand2020,vanGelder2020}, the positions in protostellar disks at which they occur (i.e., snow-lines), influences the formation and evolution of planets \citep{Oberg2011,Eistrup2016, vantHoff2017,Grassi2020}. Simultaneously, desorption processes can also shape the composition of grain surfaces. This is a consequence of the fact that all desorption mechanisms are much more efficient for volatile species (CO, O$_2$, N$_2$; \citealt{Bisschop2006,Noble2012,Cazaux2017}) than for less volatile species (H$_2$O, CH$_3$OH; \citealt{Fraser2001,Bertin2016, Cruz-Diaz2016,Martin-Domenech2016}). Since thermal desorption mechanisms alone can often not account for the diversity of species observed in star-forming regions, a further question remains open as to the extent that the prevailing physical conditions might impact ice loss, e.g. through photo-desorption, sputtering, chemical- and shock-induced processes (e.g., \citealt{Kristensen2010,Vasyunin2013,Dulieu2013,Oberg2016,Dartois2019}). Obtaining key insights on the desorption mechanisms is crucial to study the composition of ice mantles, and hence the production of complex organics, during the early phases of star-formation. 

The preferred observational approach to constrain solid-state chemistry consists of inferring abundances of solid-state molecules based on their observed gas-phase emissions \citep{Bergin2007, Oberg2009a, Whittet2011}. Although this method of indirectly deriving ice abundances from gas-phase observations is the most used to get insights into the ice composition, it relies on major assumptions, for instance on the formation pathways and the desorption efficiency of solid-state species. 

One way to test these assumptions is to combine ice and gas observations (i.e., ice- and gas-mapping techniques) and thus compare ice abundances with gas abundances towards the same region \citep{Noble2017,Perotti2020}. The evident advantage of combining the two techniques is that it enables us to study the distribution of gas-phase and solid-state molecules concurrently on the same lines of sight, and hence to address how solid-state processes are affected by physical conditions probed by gas phase-mapping such as density, temperature and radiation field gradients, turbulence and dust heating. However, the number of regions where such combined maps are available is still limited.

In this article we explore the interplay between ice and gas in the bright-rimmed cloud Barnard~35 (B35A; also known as BRC18, SFO18 and L1594) located in the $\lambda$~Orionis star-forming region \citep{Sharpless1959,Lada1976,Murdin1977,Mathieu2008,Hernandez2009, Bayo2011,Barrado2011,Kounkel2018,Ansdell2020} at a distance of 410~$\pm$~20~pc (GAIA DR2; \citealt{Zucker2019, Zucker2020}). The $\lambda$~Orionis region is characterized by a core of OB stars enclosed in a ring of dust and gas \citep{Wade1957,Heiles1974,Maddalena1987,Zhang1989, Lang2000,Dolan2002, Sahan2016}. The OB stars formed approximately 5~Myr ago, whereas the ring formation is less certain, occurring between 1 and 6~Myr ago due to a supernova explosion \citep{Dolan1999,Dolan2002,Kounkel2020}. The presence of strong stellar winds from the massive stars and ionization fronts in the region has shaped and ionized the neighboring ring of gas leading to the formation of dense molecular clouds (e.g., B35A, B30; \citealt{Barrado2018}) and photodissociation regions (PDRs; \citealt{Wolfire1989,DeVries2002,Lee2005}). For instance, the stellar wind from $\lambda$~Orionis, the most massive star of the Collinder~69 cluster (an O8III star; \citealt{Conti1974}), hits the western side of B35A compressing the cloud and forming a PDR between the molecular cloud edge and the extended HII region S264 \citep{DeVries2002,Craigon2015}. 

\begin{figure*}
\centering
\includegraphics[width=5.5in]{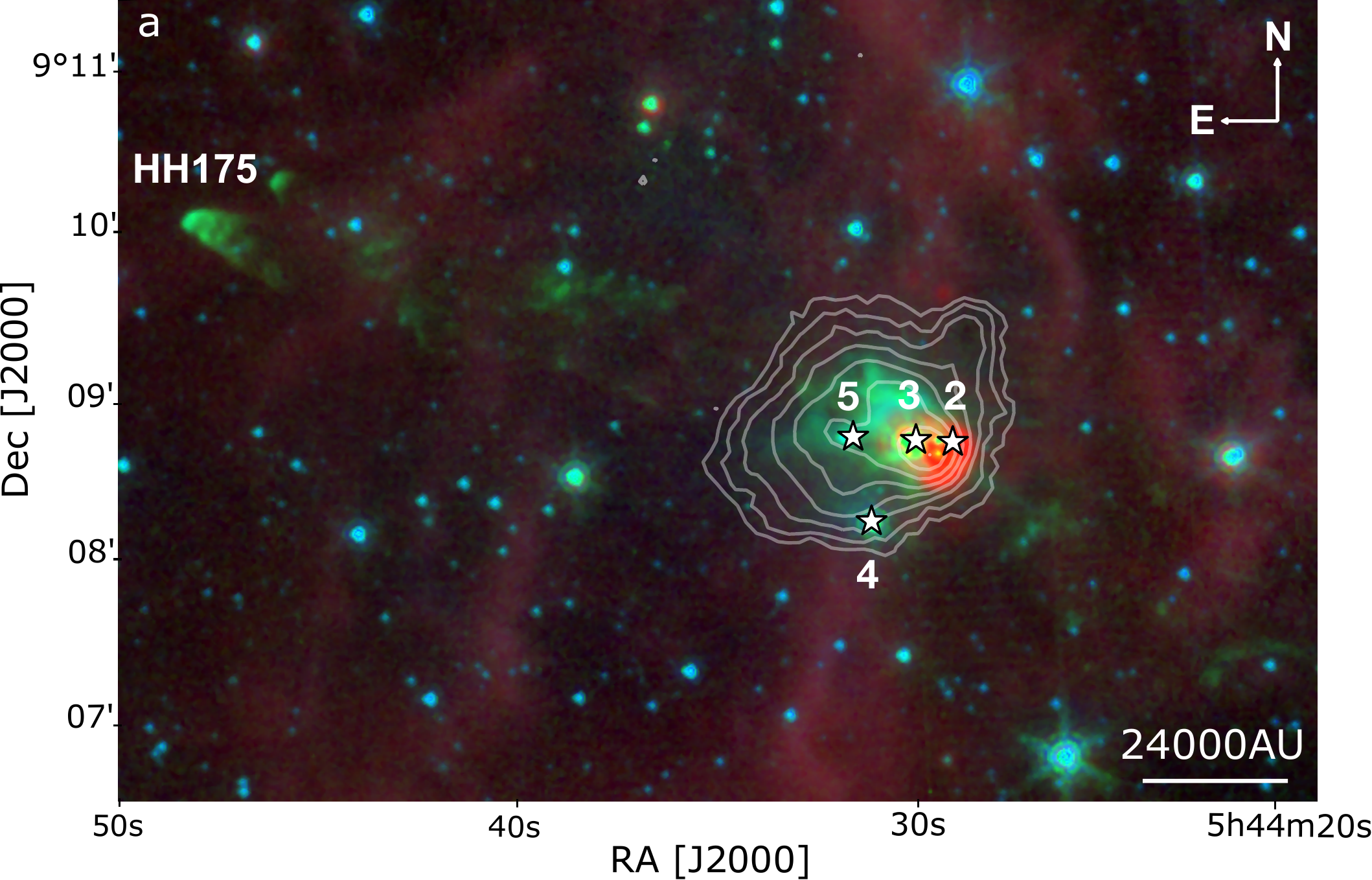}
\caption{Three-color image of B35A overlaid with SCUBA$-$2 850~$\mu$m density flux in mJy beam$^{-1}$ (\citealt{Reipurth2020}; contours are in decreasing steps of 30\% starting at 7.3~Jy beam$^{-1}$). The composite is made from \textit{Spitzer}~IRAC 3.6~$\mu$m (blue), 4.5~$\mu$m (green) and MIPS 24.0~$\mu$m (red) bands (Program ID: 139, PI: N. J.~Evans II). The white stars mark the positions of the AKARI and \textit{Spitzer} c2d identified B35A sources. \citet{Connelley2008} and \citet{Reipurth2020} show the B35A sources to be multiplet systems totalling 4 or 5 objects.}
\label{rgb_b_35a}
\end{figure*}

Star formation is currently occurring in B35A: the multiple protostellar system IRAS~05417+0907 (i.e., B35A$-$3) lies within the western side of the cloud (Figure~\ref{rgb_b_35a}).
IRAS~05417+0907 was long thought to be a single source but it is a cluster of at least four objects (B35A$-$2, B35A$-$3, B35A$-$4 and B35A$-$5) which were partially resolved by $Spitzer$ IRAC observations as part of the $Spitzer$ c2d Legacy survey\footnote{\url{https://irsa.ipac.caltech.edu/data/SPITZER/C2D/cores.html}} (\citealt{Evans2014}; Table~\ref{table:samples_sources}). B35A$-$3 (i.e., IRAS 05417+0907) is a Class~I YSO and it is the primary of a close binary system \citep{Connelley2008}; the classification of the other sources remains uncertain. B35A$-$3 is emanating a giant bipolar molecular outflow extending in the NE/SW direction relative to $\lambda$~Orionis and terminating in the Herbig-Haro object HH~175 \citep{Myers1988, Qin2003, Craigon2015, Reipurth2000}. 
A detailed multi-wavelength analysis and characterization of HH~175 and of the multiple protostellar system presented by \citet{Reipurth2020}.

The ice reservoir towards the YSOs in B35A has been extensively studied by \citet{Noble2013, Noble2017} and \citet{Suutarinen2015}. These works were based on near-infrared spectroscopic observations (2.5$-$5~$\mu$m) with the AKARI satellite. In \citet{Noble2013}, H$_2$O, CO$_2$ and CO ice features were identified towards all four YSOs, whereas \citet{Suutarinen2015} performed a multi-component, multi-line fitting of all the ice features towards the same B35A sources identified by \citet{Noble2013} and \citet{Evans2014}, and in a method similar to that employed by \citet{Perotti2020} were able to extract the H$_2$O, CO and CH$_3$OH ice column densities concurrently from the spectral data. It is these ice column densities that are used for comparison in the remainder of this paper.

In parallel, the morphology and kinematics of the gas in B35A has been investigated using single-dishes such as the \textit{James Clerck Maxwell Telescope} (JCMT) and the \textit{Institut de RadioAstronomie Millim\'etrique} (IRAM) 30~m telescope by \citet{Craigon2015}. The survey mapped $^{12}$CO, $^{13}$CO and C$^{18}$O $J$=3$-$2 and $J$=2$-$1  over a 14.6$\arcmin \times$~14.6$\arcmin$ region and confirmed the existence of a bright and dense rim along the western side of B35A, heated and compressed by the stellar wind of the nearby $\lambda$~Orionis, and of a giant bipolar outflow emanating from the YSO region.

To investigate the interaction between the solid (ice) and gas phases in the region, ice maps of B35A were compared to gas-phase maps in \citet{Noble2017}. From the combination of ice- and gas-mapping techniques, no clear correlation was found between gas or dust to ice towards B35A. The local scale variations traced by the ice-mapping were not immediately related to large scale astrophysical processes probed by the dust and gas observations.

In this paper, we present interferometric Submillimeter Array (SMA) observations of the $J_K$~=~5$_K-$4$_K$ rotational band of CH$_3$OH at 241.791~GHz and of the $J$=2$-$1 rotational bands of two CO isotopologues ($^{13}$CO and C$^{18}$O) towards the multiple protostellar system IRAS~05417+0907 in B35A. The angular resolution of the interferometric observations allows us to study the distribution of the targeted species in greater spatial detail compared to existing single-dish observations. To also recover the large-scale emission, the SMA observations are combined with single-dish data, thus resolving the protostellar system members and, concurrently probing the surrounding cloud. Furthermore, we produce ice and gas maps of a pivotal complex molecule, CH$_3$OH, and of one of its precursors, CO, to analyse the gas and ice interplay in the region. Lastly, we directly calculate gas-to-ice ratios of B35A to compare with ratios obtained for nearby star-forming regions.  

The article is organized as follows. Section~\ref{observations} describes the gas-phase observations and the archival data used to produce the gas and ice maps. Section~\ref{Results} presents the results of the observations. Section~\ref{Analysis} analyses the variations between the gas-phase and solid-state distributions of the different molecules. Section~\ref{Discussion} discusses the observational results with a particular focus on the obtained gas-to-ice ratios. Finally, Section~\ref{Conclusions} summarizes the main conclusions.  


\section{Observations and archival data}
\label{observations}

\subsection{SMA and APEX observations}
\label{SMA_APEX observations}

The sample of sources was observed on September 19, 2018 with the \textit{Submillimeter Array} (SMA; \citealt{Ho2004}). The array was in its subcompact configuration with 7 operating antennas. The targeted region was covered by two overlapping pointings; the first pointing was centered on B35A$-$3 (i.e., IRAS 05417+0907) and the second was offset by one half primary beam to the south-east. Their exact coordinates are $\alpha_{J2000} = 05^\mathrm{h}44^\mathrm{m}30^\mathrm{s}.00$, $\delta_{J2000} = +09^\circ 08\arcmin 57\farcs3$ and $\alpha_{J2000} = 05^\mathrm{h}44^\mathrm{m}30^\mathrm{s}.58$, $\delta_{J2000}=+09^\circ 08\arcmin 33\farcs8$, respectively. 

\begin{table}
\begin{center}
\caption{Sample of sources.}
\label{table:samples_sources}
\renewcommand{\arraystretch}{1.4}
\begin{tabular}{lccc} 
\hline \hline
Object$^a$  &  RA         &  DEC    & $A_V$         \\  
            & [J2000]     & [J2000]   & [mag]       \\ \hline
B35A$-$2        & 05:44:29.30 & +09:08:57.0  &   52.7$^b$       \\ 
B35A$-$3        & 05:44:30.00 & +09:08:57.3  &   54.9 $\pm$ 0.2$^c$ \\ 
B35A$-$4        & 05:44:30.85 & +09:08:26.0  &   49.6 $\pm$ 2.3$^c$ \\  
B35A$-$5        & 05:44:31.64 & +09:08:57.8  &   19.5 $\pm$ 1.5$^c$ \\ 
\hline
\end{tabular}
\end{center}
\footnotesize{\textbf{Notes.} $^a$The objects are numbered according to the ID used in \citet{Noble2013}. In \citet{Noble2017}, objects 2, 3, 4, 5 are numbered 11, 1, 6 and 12, respectively. $^b$ For a detailed description of the $A_V$ determination for B35A$-$2 see Section~\ref{H2_cd} and Appendix~\ref{appendixB}. $^c$ From the $Spitzer$~c2d catalog$^{\ref{catalog}}$.}
\end{table}

\begin{table*}
\begin{center}
\caption{Spectral data of the detected molecular transitions.}
\label{table:spectral_data_pointings}
\renewcommand{\arraystretch}{1.4}
\begin{tabular}{l c c l c c}
\hline \hline
 Transition & Frequency$^{a}$ & $A_\mathrm{ul}^{a}$ & $g_\mathrm{u}^{a}$ &  $E_\mathrm{u}^{a}$ & $n_\mathrm{cr}^{b}$ \\ 
            &  [GHz]    & [s$^{-1}$]& & [K] &  [cm$^{-3}$] \\ \hline 

C$^{18}$O $J$ = 2~$-$~1           & 219.560   &  6.01~$\times$~10$^{-7}$  &5     & 15.9 & 9.3~$\times~10^3$ \\ 
$^{13}$CO $J$ = 2~$-$~1           & 220.398   &  6.04~$\times~10^{-7}$    &5     & 15.9 & 9.4~$\times~10^3$ \\
CH$_3$OH $J$ = 5$_0~-~4_0$  E$^+$& 241.700   &  6.04~$\times~10^{-5}$    &11    & 47.9  & 5.5~$\times~10^5$ \\
CH$_3$OH $J$ = 5$_1~-~4_1$  E$^-$& 241.767   &  5.81~$\times~10^{-5}$    &11    & 40.4  & 4.8~$\times~10^5$ \\
CH$_3$OH $J$ = 5$_0~-~4_0$  A$^+$& 241.791   &  6.05~$\times~10^{-5}$    &11    & 34.8  & 5.0~$\times~10^5$ \\
CH$_3$OH $J$ =  5$_1~-~4_1$  E$^+$& 241.879   &  5.96~$\times~10^{-5}$   &11   & 55.9   & 4.9~$\times~10^5$ \\
CH$_3$OH $J$ = 5$_2~-~4_2$  E$^-$& 241.904   &  5.09~$\times~10^{-5}$    &11   & 60.7   & 4.2~$\times~10^5$ \\
\hline 
\end{tabular}
\end{center}

\begin{center}
\begin{tablenotes}
\small
\item{\textbf{Notes.}$^{a}$ From the Cologne Database for Molecular Spectroscopy (CDMS; \citet{Muller2001}) and the Jet Propulsion Laboratory catalog \citep{Pickett1998}. $^{b}$ Calculated using a collisional temperature of 20~K and collisional rates from the Leiden Atomic and Molecular Database (LAMDA; \citealt{Schoier2005}). The references for the collisional rates are \citet{Yang2010} for the CO isotopologues and \citet{Rabli2010} for CH$_3$OH.}
\end{tablenotes}
\end{center}
\end{table*}

The SMA observations probed frequencies spanning from 214.3 to 245.6~GHz with a spectral resolution of 0.1~MHz (0.135~km~s$^{-1}$). Among other species this frequency range covers the CH$_3$OH $J_K$~=~5$_K-$4$_K$ branch at 241.791~GHz, $^{13}$CO $J$~=~2$-$1 at 220.398~GHz and C$^{18}$O $J$~=~2$-$1 at 219.560~GHz (Table~\ref{table:spectral_data_pointings}). The observational setup consisted of a first long integration on the bandpass calibrator, the strong quasar 3c84, followed by alternated integrations on the source and on the gain calibrator, the quasar J0510+180. The absolute flux scale was obtained through observations of Uranus. The SMA data set was both calibrated and imaged using CASA\footnote{\url{http://casa.nrao.edu/}} \citep{McMullin2007}. At these frequencies, the typical SMA beam-sizes were 7$\farcs$3~$\times$~2$\farcs$4 with a position angle of -32.2$^{\circ}$ for CH$_3$OH, and 7$\farcs$9~$\times$~2$\farcs$6 with a position angle of -31.2$^{\circ}$ for $^{13}$CO and C$^{18}$O.

To trace the more extended structures in the region, the SMA data were complemented by maps obtained using the \textit{Atacama Pathfinder EXperiment} (APEX; \citealt{Gusten2006}) on August 20$-$22, 2018. The single-dish observations covered frequencies between 236$-$243.8~GHz, matching the SMA 240~GHz receiver upper side band. The spectral resolution of the APEX observations was 0.061~MHz (0.076~km~s$^{-1}$). The APEX map size was 100$''~\times$~125$''$ and it extended over both SMA primary beams. The coordinates of the APEX pointing are $\alpha_{J2000} = 05^\mathrm{h}44^\mathrm{m}30^\mathrm{s}.00$, $\delta_{J2000} = +09^\circ 08\arcmin 57\farcs3$. The APEX beam size was 27$\farcs$4 for the CH$_3$OH $J_K$~=~5$_K-$4$_K$ lines emission. The APEX data set was reduced using the GILDAS package CLASS\footnote{\url{http://www.iram.fr/IRAMFR/GILDAS}}. 
At a later stage, the reduced APEX data cube was imported to CASA and combined with the interferometric data using the feathering technique. A description of the combination procedure is given in Appendix~\ref{data_combination}. 

\begin{figure*}
  \centering
  \includegraphics[trim={0 0 0 0}, clip, width=7in]{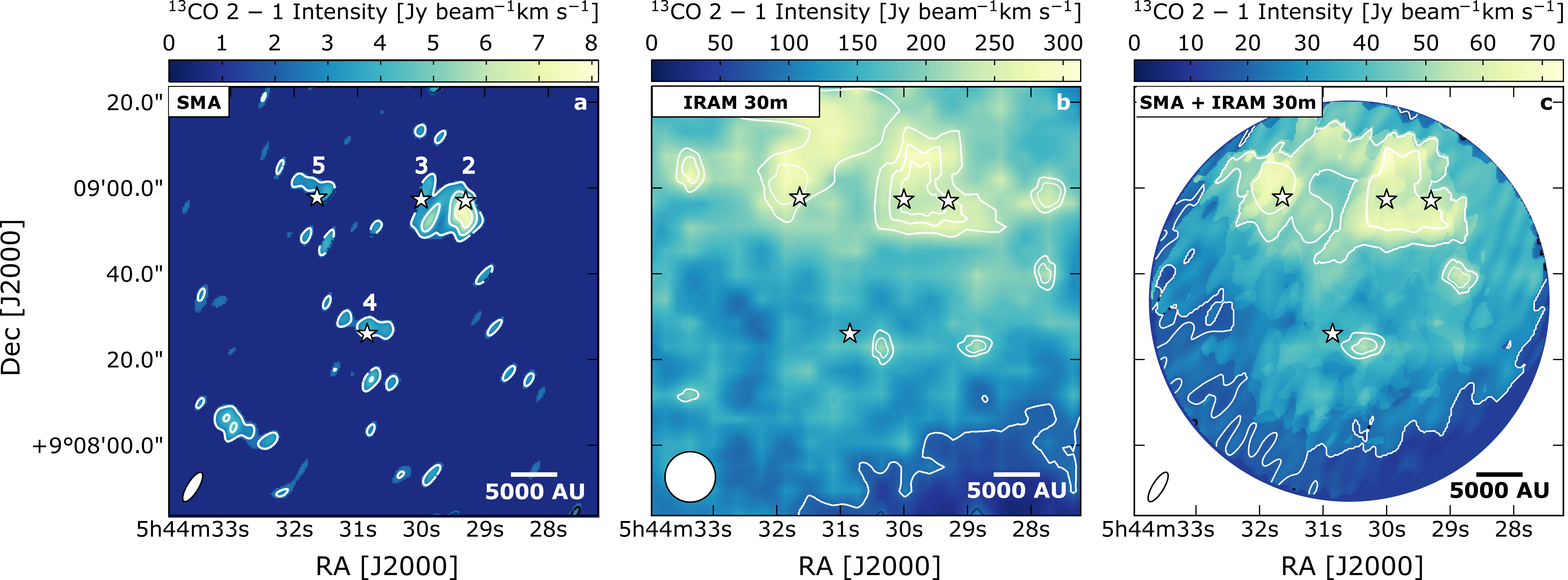}
      \caption{Integrated intensity maps for $^{13}$CO $J$=2$-$1 detected by the SMA (a), by the IRAM~30~m telescope (b), and in the combined interferometric SMA and single-dish IRAM~30~m data (c). All lines are integrated between 7 and 16~km~s$^{-1}$. Contours are at 5$\sigma$, 10$\sigma$, 15$\sigma$, etc.  ($\sigma_\mathrm{SMA}$= 0.42~Jy beam$^{-1}$km~s$^{-1}$, $\sigma_\mathrm{IRAM~30~m}$= 6~Jy beam$^{-1}$km~s$^{-1}$, $\sigma_\mathrm{SMA+IRAM~30~m}$= 1.03~Jy beam$^{-1}$km~s$^{-1}$). In panel (c), the white area outlines the primary beam of the SMA observations. The synthesized beams are displayed in white in the bottom left corner of each panel. The white stars mark the position of the targeted B35A sources.}
         \label{mom0_13co}
\end{figure*}

\subsection{JCMT/SCUBA-2, Spitzer IRAC, 2MASS, IRAM~30~m and AKARI data} 
\label{JCMT/SCUBA-2 and AKARI data}
Ancillary data to the SMA and APEX observations were adopted in this study. To construct H$_2$ column density maps of B35A, we used submillimeter continuum maps at 850~$\mu$m obtained with the SCUBA$-$2 camera at the \textit{James Clerk Maxwell Telescope} (JCMT) by \citet{Reipurth2020}, visual extinction values from the $Spitzer$ c2d catalog\footnote{\label{catalog}\url{https://irsa.ipac.caltech.edu/data/SPITZER/C2D/cores.html}} and 2MASS and $Spitzer$ IRAC photometric data \citep{Skrutskie2006,Evans2014}. \textit{Institut de RadioAstronomie Millim\'etrique} (IRAM) 30~m telescope observations \citep{Craigon2015} were used to trace the extended emission for $^{13}$CO and C$^{18}$O $J$~=~2$-$1. Finally, to produce gas-ice maps of B35A we made use of the ice column densities determined by \citet{Suutarinen2015}, from \textit{AKARI} satellite observations.

\section{Results}
\label{Results}
This section lists the observational results, supplying a summary of the methods employed to determine CO and CH$_3$OH gas column densities (Section~\ref{Gas_cd} and Appendix~\ref{appendixA}). Additionally, it presents the H$_2$O, CO and CH$_3$OH ice column densities (Section~\ref{Ice_cd}) and the two H$_2$ column density maps used in the calculation of the abundances of the ice and gas species (Sect.~\ref{H2_cd}). 

\subsection{Gas-phase species}
\label{Gas_cd}

\begin{figure*}
  \centering
   \includegraphics[trim={0 0 0 0}, clip, width=7in]{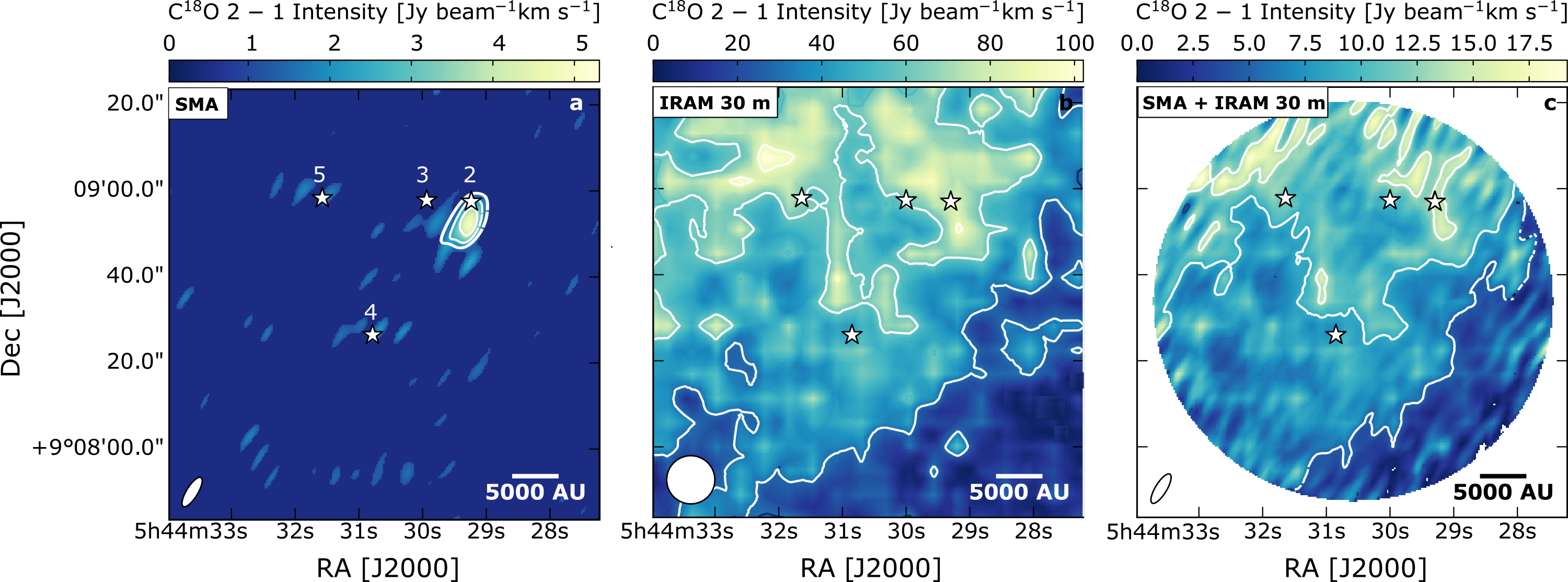}
      \caption{Integrated intensity maps for C$^{18}$O $J$=2 $-$ detected by the SMA (a), by the IRAM~30~m telescope (b), and in the combined interferometric SMA and single-dish IRAM~30~m data (c). All lines are integrated between 10 and 15~km~s$^{-1}$. Contours are at 5$\sigma$, 10$\sigma$, 15$\sigma$ ($\sigma_\mathrm{SMA}$= 0.37~Jy beam$^{-1}$km~s$^{-1}$, $\sigma_\mathrm{IRAM~30~m}$= 5~Jy beam$^{-1}$km~s$^{-1}$, $\sigma_\mathrm{SMA+IRAM~30~m}$= 0.40~Jy beam$^{-1}$km~s$^{-1}$). 
      In panel (c), the white area outlines the primary beam of the SMA observations.
      The synthesized beams are displayed in white in the bottom left corner of each panel. The white stars mark the position of the targeted B35A sources.}
         \label{mom0_c18o}
\end{figure*}

\begin{figure*}
  \centering
  \includegraphics[trim={0 0 0 0}, clip, width=7in]{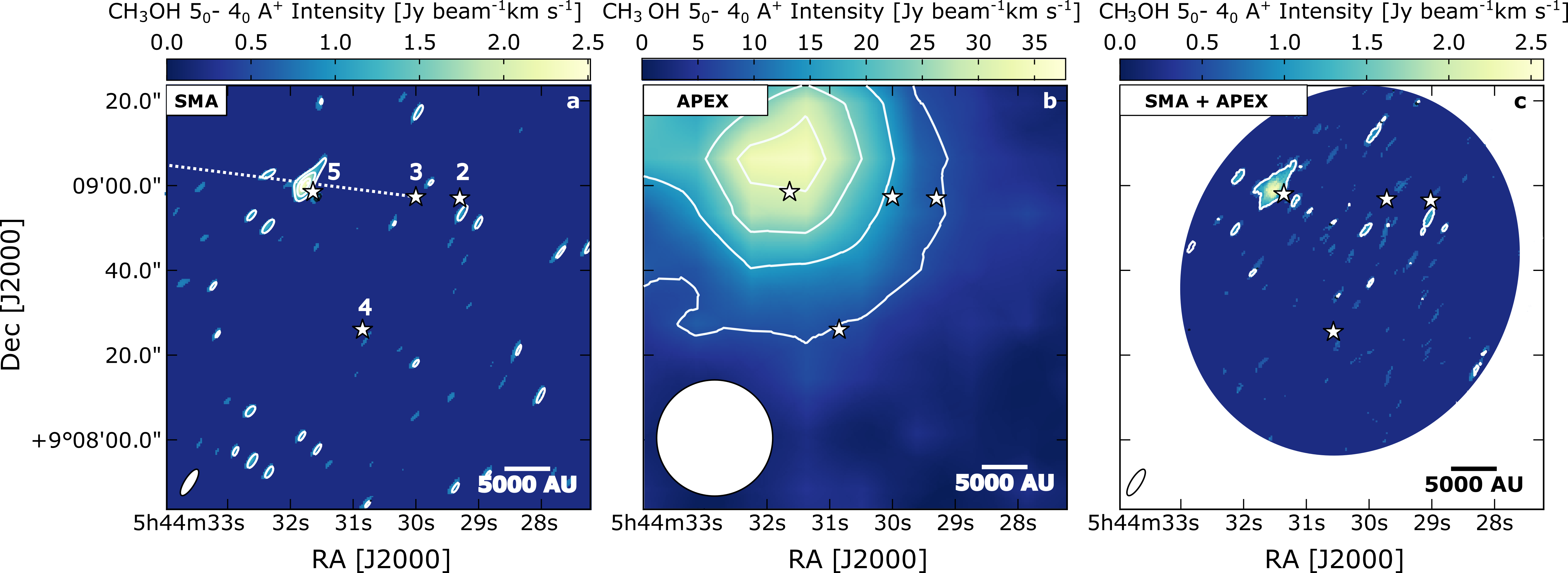}
      \caption{Integrated intensity maps for CH$_3$OH $J$=5$_0$ $-$ 4~$_0$~A$^+$ detected by the SMA (a), by the APEX telescope (b), and in the combined interferometric SMA and single-dish APEX data (c). All lines are integrated between 6 and 12.5~km~s$^{-1}$. Contours start at 5$\sigma$ ($\sigma_\mathrm{SMA}$= 0.04~Jy beam$^{-1}$km~s$^{-1}$, $\sigma_\mathrm{APEX}$= 4~Jy beam$^{-1}$km~s$^{-1}$, $\sigma_\mathrm{SMA+APEX}$= 0.04~Jy beam$^{-1}$km~s$^{-1}$) and follow a step of 5$\sigma$. In panel (c), the white area outlines the primary beam of the SMA observations. The synthesized beams are displayed in white in the bottom left corner of each panel. The white stars mark the position of the targeted B35A sources and the blue dotted line indicates the direction of the giant outflow terminating in HH~175.}
         \label{mom0_meth}
\end{figure*}

\begin{table*}
\begin{flushleft}
\caption{Total ice and gas column densities towards the B35A sources.}
\label{table:summary_cd}
\renewcommand{\arraystretch}{2.2}
\scalebox{0.95}{
\begin{tabular}{lcccccccc} 
\hline \hline
Object & $N^\mathrm{ice}_\mathrm{H_2O\ tot}$ & $N^\mathrm{ice}_\mathrm{CO \ tot}$ & $N^\mathrm{ice}_\mathrm{CH_3OH \ tot}$ & $N^\mathrm{gas}_\mathrm{C^{18}O \ tot}$ &  $N^\mathrm{gas}_\mathrm{^{12}CO \ tot}$ & $N^\mathrm{gas}_\mathrm{CH_3OH \ tot}$  & $^{\dagger}N_\mathrm{H_2}^\mathrm{SCUBA-2}$ & $N_\mathrm{H_2}^{A_V}$ \\  
      &[10$^{18}$cm$^{-2}$]&  [10$^{18}$cm$^{-2}$] &[10$^{18}$cm$^{-2}$] & [10$^{15}$cm$^{-2}$] & [10$^{18}$cm$^{-2}$] & [10$^{14}$cm$^{-2}$]& [10$^{22}$cm$^{-2}$] & [10$^{22}$cm$^{-2}$] \\ \hline 
B35A$-$2   & 2.96 $\pm$ 0.13 & 0.34 $\pm$ 0.03 & 0.26 $\pm$ 0.03 & 9.38 $\pm$ 1.92 & 5.22 $\pm$ 1.07 & ...  & 1.76 $\pm$ 0.09 & 7.22$^{*}$      \\  
B35A$-$3   & 2.96 $\pm$ 0.12 & 0.32 $\pm$ 0.03 & 0.16 $\pm$ 0.03 & 7.02 $\pm$ 1.44 & 3.91 $\pm$ 0.80 & ...  & 1.69 $\pm$ 0.08 & 7.52 $\pm$ 0.02 \\  
B35A$-$4   & 3.35 $\pm$ 0.11 & 0.28 $\pm$ 0.06 & 0.10 $\pm$ 0.06 & 4.99 $\pm$ 1.03 & 2.78 $\pm$ 0.57 & ...  & 0.39 $\pm$ 0.02 & 6.79 $\pm$ 0.68 \\  
B35A$-$5   & 0.78 $\pm$ 0.12 &  < 0.18  & < 0.18 & 7.11 $\pm$ 1.46 & 3.96 $\pm$ 0.81  & 1.91 $\pm$ 0.24 & 1.09 $\pm$ 0.05 & 2.66 $\pm$ 0.20  \\  
\hline 
\end{tabular}
}
\end{flushleft}
\footnotesize{\textbf{Notes.} Columns 2$-$4 display the ice column densities from \citet{Suutarinen2015}. Columns 5$-$7 list the gas column densities for $T_\mathrm{rot}$=~25~K. The errors are estimated based on the rms noise of the spectra and on the $\sim$20\% calibration uncertainty. Non-detections are indicated by "...". $^{\dagger}$~$N_\mathrm{H_2}^\mathrm{SCUBA-2}$ is calculated using $T_\mathrm{dust}$=~25~K. The errors are calculated from the 5\% flux calibration uncertainty and do not take into account the uncertainty on the $T_\mathrm{dust}$.} $^{*}$ The uncertainty on the $A_V$ and consequently on the $N_\mathrm{H_2}^{A_V}$ for B35$-$2 is not estimated because this object is not detected in the J, H and K bands.
\end{table*}

The spectral line data of the detected molecular transitions are listed in Table~\ref{table:spectral_data_pointings}. Figures \ref{mom0_13co}--\ref{mom0_meth} display moment~0 maps of the $^{13}$CO and C$^{18}$O $J$=2$-$1 lines and CH$_3$OH $J$=5$_0-4_0$ A$^+$ line, using SMA, IRAM~30~m and APEX data. Moment 0 maps of the CH$_3$OH $J$=5$_0-4_0$ A$^+$ line are presented in this section as this transition is the brightest of the CH$_3$OH $J$=5$_K-4_K$ branch at 241.791 GHz. The maps show that the SMA observations filter out spatially extended emission related to the B35A cloud. The SMA data only recover $\approx$10\% of the extended emission detected in the single-dish data. Thus, consequently we need to combine the interferometric data with the single-dish maps in order not to underestimate the column densities severely.

Panels~(a) of Figures \ref{mom0_13co}--\ref{mom0_c18o} show that the interferometric emission is predominantly compact and the peak intensity is seen, for both CO isotopologues, at the location of B35A$-$2. The emission observed in the IRAM~30~m data sets (panels~(b) of Figures \ref{mom0_13co}--\ref{mom0_c18o}) is extended and mostly concentrated around B35A$-$2, B35A$-$3 and B35A$-$5. In the combined SMA~+~IRAM~30~m maps (panels~(c) of Fig. \ref{mom0_13co}--\ref{mom0_c18o}) the peak emission is also located in the region where B35A$-$2, B35A$-$3 and B35A$-$5 are present.

In contrast, the observed CH$_3$OH peak intensity is exclusively localized at the B35A$-$5 position in the SMA data (panel~(a) of Fig.\ref{mom0_meth}). B35A$-$5 lies along the eastern lobe of the giant bipolar molecular outflow emanated from IRAS~05417+0907 (i.e., B35A$-$3) and terminating at the location of the Herbig-Haro object HH~175 \citep{Craigon2015}. The CH$_3$OH emission is extended in the APEX data (panel~(b) of Fig.\ref{mom0_meth}) and hence not resolved at one particular source position. The emission in the SMA~+~APEX moment 0 map (panel~(c) of Fig.\ref{mom0_meth}) is compact and concentrated in one ridge in the proximity of B35A$-$5. The channel maps (Figures~\ref{ch_maps_13co}$-$\ref{ch_maps_ch3oh}) and the spectra (Figure~\ref{spectra}) extracted from the combined interferometric and single-dish data sets show predominantly blue-shifted components (Appendix~\ref{channel maps and spectra}). Blue- and red-shifted components are seen for the CO isotopologues, which were previously observed by \citet{DeVries2002}, \citet{Craigon2015} and \citet{Reipurth2020} and attributed to outflowing gas.

The $^{13}$CO emission is optically thick towards the B35A sources, therefore the $^{13}$CO column densities are underestimated towards the targeted region (Appendix~\ref{otically_thick_emission}). Consequently, the optically thin C$^{18}$O emission is adopted to estimate the column density of $^{12}$CO in B35A. First, C$^{18}$O column densities towards the protostellar system members are obtained from the integrated line intensities of the combined SMA~+~IRAM~30~m maps (panels (c) of Figure~\ref{mom0_c18o}), assuming optically thin emission and a kinetic temperature for the YSO region equal to 25~K \citep{Craigon2015}. The formalism adopted in the column density calculation is presented in Appendix~\ref{gas-phase_cd_formalism}, whereas the calculated C$^{18}$O column densities and their uncertainties are listed in Table~\ref{table:summary_cd}. The C$^{18}$O column densities are converted to $^{12}$CO column densities assuming a $^{16}$O/$^{18}$O isotope ratio of 557~$\pm$~30 \citep{Wilson1999}. The resulting $^{12}$CO column densities (Table~\ref{table:summary_cd}) are in good agreement with the values reported in \citet{Craigon2015}. 

The CH$_3$OH column density towards B35A$-$5 is estimated from the integrated line intensities of the combined SMA~+~APEX maps for the five CH$_3$OH lines (Table~\ref{table:ch3oh_intensities}), assuming optically thin CH$_3$OH emission, a kinetic temperature of 25~K \citep{Craigon2015} and local thermodynamic equilibrium (LTE) conditions. The CH$_3$OH column density and its uncertainty is reported in Table~\ref{table:summary_cd}. For all the gas-phase species, the uncertainty on the column densities is estimated based on the spectral rms noise and on the $\sim$20\% calibration uncertainty.

\subsection{Ice column densities}
\label{Ice_cd}

The ice column densities of H$_2$O, CO and CH$_3$OH made use of in this paper were originally derived by \citet{Suutarinen2015} from the AKARI (2.5$-$5~$\mu$m) NIR spectra of the targeted B35A sources. \citet{Suutarinen2015} adopted a non-heuristic approach, employing ice laboratory data and analytical functions to concurrently account for the contribution of a number of molecules to the observed ice band features. A detailed description of the AKARI observations is given in \citet{Noble2013} and the methodology adopted to derive the column densities of the major ice species can be found in \citet{Suutarinen2015}.

The total ice column densities from \citet{Suutarinen2015} are listed in Table~\ref{table:summary_cd}. Among the three molecules, H$_2$O ice is the most abundant in the ices of B35A, with column densities up to 3.35~$\times$~10$^{18}$~cm$^{-2}$, followed by CO and CH$_3$OH (Table~\ref{table:summary_cd}). The column densities of the latter two molecules are comparatively one order of magnitude less abundant than H$_2$O ice. Upper limits are given for the CO and CH$_3$OH column densities towards B35A$-$5 because of difficulties in distinguishing the absorption features in the spectrum of this YSO \citep{Noble2013,Suutarinen2015}. A trend is observed for the CO and CH$_3$OH values: the column densities towards B35A$-$2 are the highest reported, followed by B35A$-$3, B35A$-$4 and B35A$-$5. The CO and CH$_3$OH ice column densities appear to follow the visual extinction with B35A$-$2 and B35A$-$3 being the most extincted sources and showing the highest CO and CH$_3$OH column densities (Tables~\ref{table:samples_sources} \& \ref{table:summary_cd}). In this regard, it is important to recall that B35A$-$2 is not detected in the J, H and K bands, making the determination of its visual extinction uncertain (Section~\ref{H2_cd} and Appendix~\ref{appendixB}). 

\subsection{H$_2$ column densities}
\label{H2_cd}
An important factor to consider when combining ice- and gas-mapping techniques is that ice absorption and dust and gas emissions may probe different spatial scales \citep{Noble2017}. As a matter of fact, the targeted sources may be embedded to different depths in the B35A cloud and consequently, the gas and ice observations may not be tracing the same columns of material. Therefore, the search for gas-ice correlations towards B35A has to be performed by comparing gas and ice abundances relative to H$_2$, following the approach adopted in \citet{Perotti2020}. 

\begin{figure}[b!]
\centering
\includegraphics[trim={50 50 10 75},clip,width=3.6in]{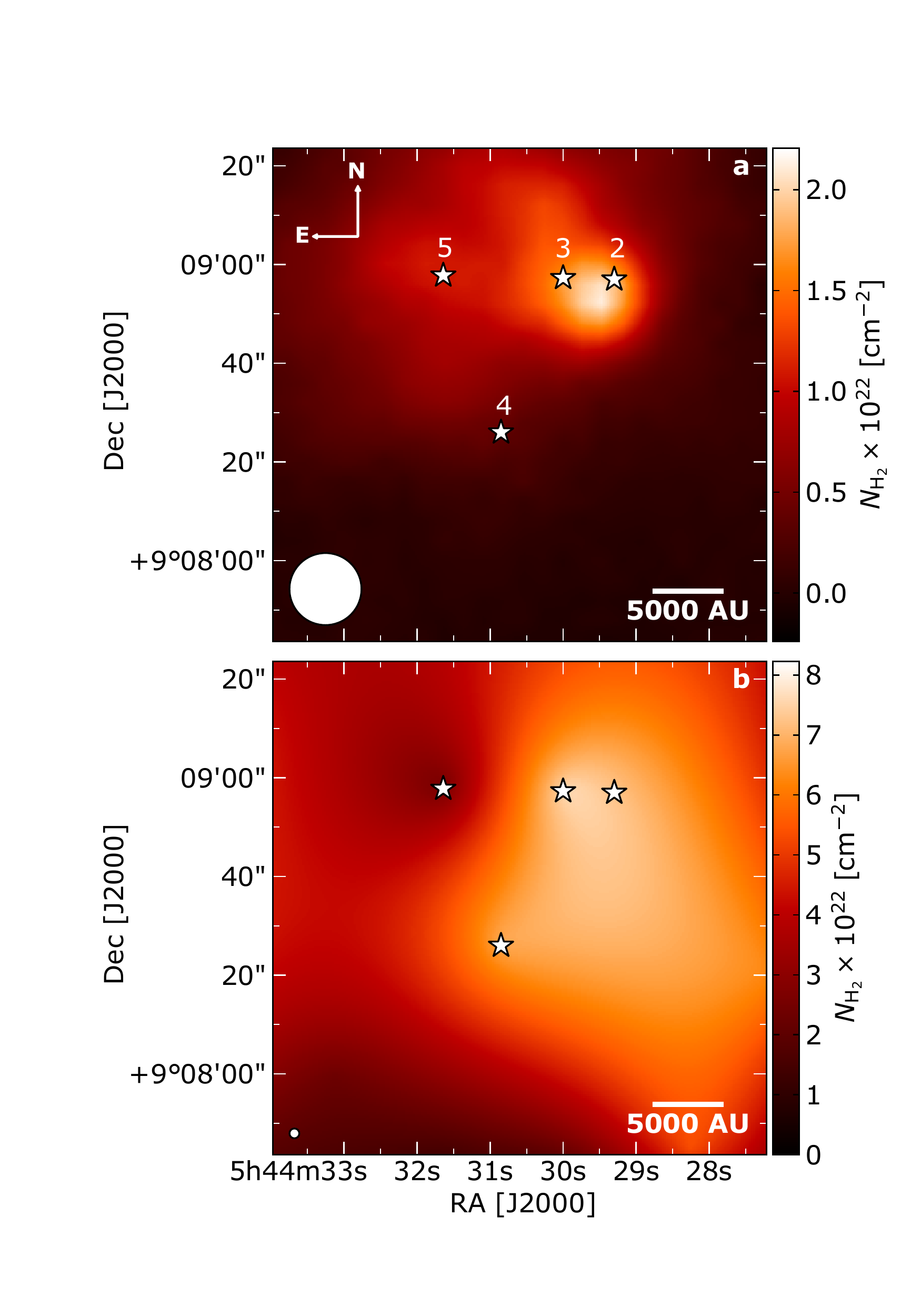}
\caption{H$_2$ column density maps of B35A. $a$: $N_\mathrm{H_2}$ map calculated from SCUBA$-$2 dust emission at 850~$\mu$m \citep{Reipurth2020}. $b$: $N_\mathrm{H_2}$ map calculated from the visual extinction. The synthesized beams are displayed in white in the bottom left corner of each panel. The white stars mark the positions of the targeted B35A sources.}
\label{NH2_maps}
\end{figure}

To keep the gas and ice observations in their own reference frame, two H$_2$ column density maps are produced: one to derive gas abundances and one to determine ice abundances (Figure~\ref{NH2_maps}). For the gas observations, estimates of the H$_2$ column density are made using submillimeter continuum maps of B35A at 850$\mu$m (SCUBA$-$2; \citealt{Reipurth2020}), under the assumption that the continuum emission is originated from optically thin thermal dust radiation \citep{Kauffmann2008}. In this regime, the strength of the submillimeter radiation is dependent on the column density ($N$), the opacity ($\kappa_\nu$) and the dust temperature ($T$). The adopted value for the opacity per unit dust+gas mass at 850~$\mu$m is 0.0182~cm$^2$ g$^{-1}$ ("OH5 dust"; \citealt{Ossenkopf1994}). The temperature of B35A has been estimated by \citet{Morgan2008} and \citet{Craigon2015}. Both studies found two distinct regimes within the cloud: a cold region ($T_\mathrm{gas}=$~10~–~20~K, $T_\mathrm{dust}= $ 18~K) in the shielded cloud interior to the east of the YSOs and a warm region ($T_\mathrm{gas}=$~20~–~30~K) to the west of the YSOs. Since the region where the YSOs are located lies close to the warm cloud western edge, a $T_\mathrm{dust}= $ 25~K is adopted to estimate the H$_2$ column density map illustrated in Figure~\ref{NH2_maps}~(a) based on \citet{Craigon2015} and \citet{Reipurth2020}. The H$_2$ column density towards the B35A sources is reported in Table~\ref{table:summary_cd}. The error on the H$_2$ column density was estimated according to the 5\% flux calibration uncertainty \citep{Dempsey2013}. It does not include the errors on $T_\mathrm{dust}$ due to the uncertainties on the upper and lower limits for $T_\mathrm{dust}$ in the YSOs region. 

Increasing the dust temperature to 30~K would lower the H$_2$ column density by 22\% and consequently increase the abundance of the gas-phase species by the same amount – while lowering the dust temperature to 18~K would increase the H$_2$ column density by 61\% and consequently lowering the gas abundance likewise. The above estimates assume an excitation temperature equal to 25~K for CO and CH$_3$OH. Increasing both the excitation temperature and the dust temperature to 30~K would result in gas abundances increasing by 24\%. Conversely, lowering both temperatures to 18~K, would lower the abundances by 33\%.
A comprehensive description of the production of the H$_2$ column density map from SCUBA$-$2 measurements is given in Appendix~C of \citet{Perotti2020}. 

The H$_2$ column density map calculated from the SCUBA$-$2 measurements supplies an accurate estimate of the total beam-averaged amount of gas towards B35A, and therefore provides a useful reference for the optically thin gas-phase tracers. However, the derived H$_2$ column density map can not directly be related to the AKARI data that supply (pencil-beam) measurements of the column densities towards the infrared sources that may or may not be embedded within the B35A cloud. 

For the ice observations, the H$_2$ column density map is therefore obtained by performing a linear interpolation of the tabulated visual extinction ($A_V$) values for B35A taken from the $Spitzer$ c2d catalog\footnote{\url{https://irsa.ipac.caltech.edu/data/SPITZER/C2D/cores.html}} (see Appendix~\ref{appendixB}). Since no calculated $A_V$ values are reported for B35A$-$2 and B35A$-$3, the $A_V$ for B35A$-$3 is acquired by de-reddening the spectral energy distribution (SED) at the 2MASS J, H, K photometric points to fit a blackbody and using a second blackbody to model the infrared excess at longer wavelengths. A detailed description of the fitting procedure adopted for B35A$-$3 is given in Appendix~\ref{appendixB}. Following \citet{Evans2009} and \citet{Chapman2009} we adopt an extinction law for dense interstellar medium gas with $R_V =$~5.5 from \citet{Weingartner2001} to calculate $A_V$. The visual extinction for B35A$-$2 could not be estimated through a similar procedure due to the lack of near-IR photometry data available for this object, thus the $A_V$ for this source is obtained through the interpolation of the available visual extinction measurements for B35A. The final $A_V$ values of the four sources are listed in Table~\ref{table:samples_sources}. Finally, the visual extinction is converted to $H_2$ column density using the relation: $N_\mathrm{H_2}= 1.37 \times 10^{21}$~cm$^{-2}$~($A_V$/mag) established for dense interstellar medium gas \citep{Evans2009}. The H$_2$ column density calculated from the $A_V$ map is shown in Figure~\ref{NH2_maps}~(b). 

The H$_2$ column densities calculated from both methods are of the same order of magnitude,  10$^{22}$~cm$^{-2}$ (Table~\ref{table:summary_cd}). The column densities differ by approximately a factor of 4 possibly due to variations in the exact column densities traced by SCUBA$-$2 and $A_V$ measurements and assumptions on the dust temperature.
In both H$_2$ column density maps (Fig.~\ref{NH2_maps}~a and b), it is seen that the four sources lie in the densest region of the cloud, confirming the results previously presented in \citet{Craigon2015} and \citet{Reipurth2020}. Although the two maps were calculated using two different methods, they reproduce the same trend: B35A$-$2 and B35A$-$3 are located where the H$_2$ column density peaks, whereas B35A$-$4 and B35A$-$5 are situated in less dense regions. However, in the H$_2$ column density map calculated from the visual extinction (Fig.~\ref{NH2_maps}~b), lower $N_\mathrm{H_2}$ values are reported for B35A$-$5 with respect to the other YSOs, whereas this is not the case in the H$_2$ column density map calculated from the SCUBA$-$2 map (Fig.~\ref{NH2_maps}~a) where lower $N_\mathrm{H_2}$ values are observed for B35A$-$4. This difference can be explained by recalling that B35A$-$5 is the less extincted YSO ($A_V$~= 19.5~mag) and it is located along the trajectory of the giant outflow launched by B35A$-$3. Consequently, the SCUBA$-$2 observations (14.6$\arcsec$~beam) are likely tracing the more extended structure of the dust emission in the region and not resolving the more compact emission towards each individual objects. It is worth observing that embedded protostars and Herbig-Haro objects can substantially affect the dust emission morphology \citep{Chandler1996}, hence, the $N_\mathrm{H_2}$ enhancement seen in Figure~\ref{NH2_maps}~(a) likely reflects dust heating by the embedded protostars and by HH175 resulting in an increase of the submillimeter continuum flux, rather than in an higher column density.



\section{Analysis}
\label{Analysis}

\subsection{Gas-ice maps}
\label{comb_maps}

\begin{figure}
\includegraphics[trim={0 0 0 0},clip,width=3.2in]{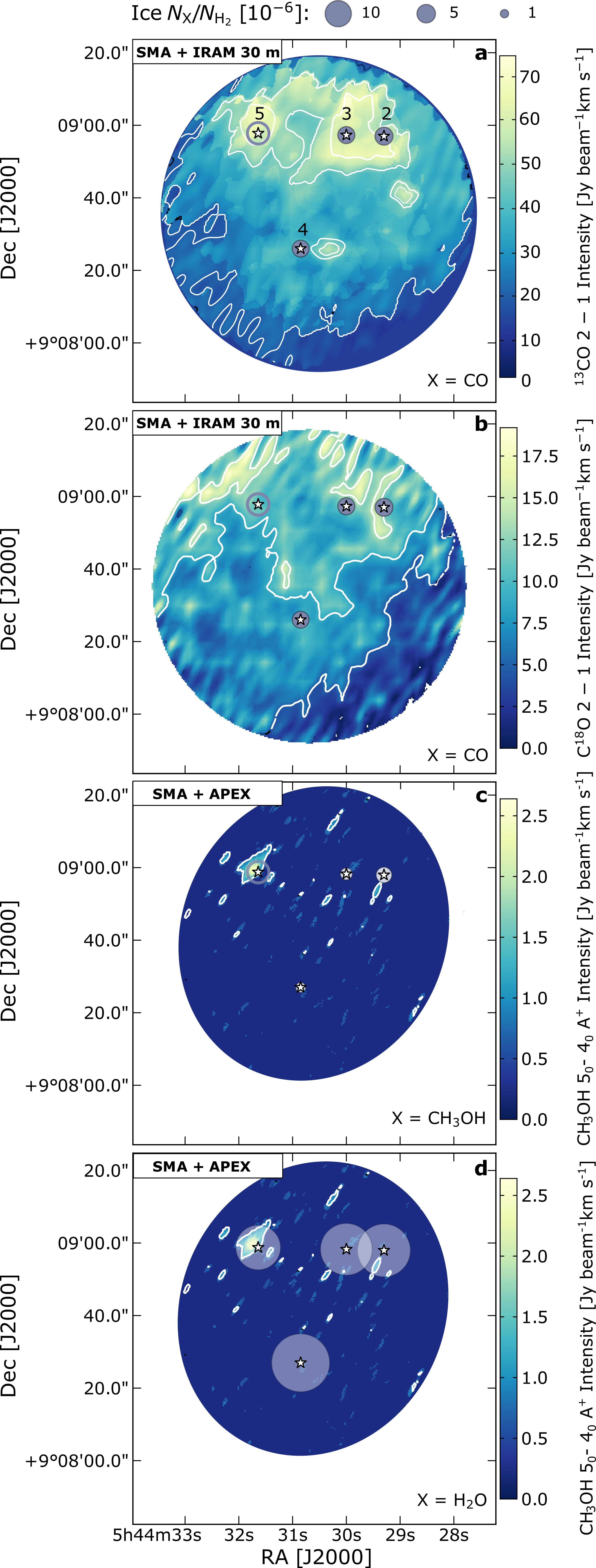}
\caption{Gas-ice maps of B35A. Ice abundances are indicated as filled grey (panels a,b) or white circles (panels c and d), upper limits are displayed as empty circles.
Contour levels are 5$\sigma$, 10$\sigma$, 15$\sigma$. \textit{a:}~CO ice abundances on gas $^{13}$CO~2$-$1; \textit{b:}~CO ice abundances on gas C$^{18}$O~2$-$1 ; \textit{c:}~CH$_3$OH ice abundances on gas CH$_3$OH~5$_0-4_0$~A$^+$. \textit{d:}~H$_2$O ice abundances
on gas CH$_3$OH~5$_0-4_0$~A$^+$. The white area outlines the primary beam of the SMA observations. The white stars mark the position of the targeted B35A sources.}
\label{ig_maps}
\end{figure}

In Figure~\ref{ig_maps}, the distributions of gas-phase $^{13}$CO 2$-$1 (panel~a) and C$^{18}$O 2$-$1 (panel~b) emissions are compared to CO ice abundances. Both CO isotopologue emissions are concentrated at the B35A$-$2, B35A$-$3, B35A$-$5 source positions. The emission towards these three sources is of comparable intensity but it drops in the surroundings of B35A$-$4 and especially towards the southern edge of the cloud. The CO ice abundances with respect to H$_2$ are not characterized by large variations, instead, they are quite uniform and consistent within the uncertainties (Table~\ref{table:summary_abund}). Only an upper limit could be determined for the CO ice column density towards B35A$-$5, due to the uncertainty in distinguishing the absorption feature at 3.53~$\mu$m in the spectrum of this object.

\begin{table*}
\begin{center}
\caption{Ice and gas abundances relative to H$_2$ towards the B35A sources.}
\label{table:summary_abund}
\renewcommand{\arraystretch}{1.8}
\begin{tabular}{lccccc} 
\hline \hline
Object & $X^\mathrm{ice}_\mathrm{H_2O}$ & $X^\mathrm{ice}_\mathrm{CO}$ & $X^\mathrm{ice}_\mathrm{CH_3OH}$ & $X^\mathrm{gas}_\mathrm{^{12}CO}$ & $X^\mathrm{gas}_\mathrm{CH_3OH}$ \\  
      &[10$^{-5}$]&  [10$^{-5}$] &[10$^{-5}$] & [10$^{-4}$] & [10$^{-8}$] \\ \hline 
B35A$-$2   & 4.10 $\pm$ 0.18 & 0.47 $\pm$ 0.04 & 0.36   $\pm$ 0.04 & 2.97 $\pm$ 0.63 & ...  \\  
B35A$-$3   & 3.99 $\pm$ 0.16 & 0.43 $\pm$ 0.04 & 0.21   $\pm$ 0.04 & 2.31 $\pm$ 0.49 & ...  \\  
B35A$-$4   & 4.93 $\pm$ 0.52 & 0.41 $\pm$ 0.50 & 0.15   $\pm$ 0.09 & 7.13 $\pm$ 1.50 & ...  \\  
B35A$-$5   & 2.93 $\pm$ 0.50 &  < 0.68  & < 0.68 & 3.63 $\pm$ 0.76 & 1.75 $\pm$ 0.24 \\  
\hline 
\end{tabular}
\end{center}
\footnotesize{\textbf{Notes.} The column densities employed in the determination of the ice and gas abundances of H$_2$O, CO and CH$_3$OH are listed in Table~\ref{table:summary_cd}. The abundances of the gaseous species are obtained using $N_\mathrm{H_2}^\mathrm{SCUBA-2}$, whereas the abundances of the ice species are relative to $N_\mathrm{H_2}^{A_V}$ (Table~\ref{table:summary_cd}). "..." mark the non-detections of gas-phase CH$_3$OH.}
\end{table*}

\begin{figure*}
\centering
\includegraphics[width=7.35in]{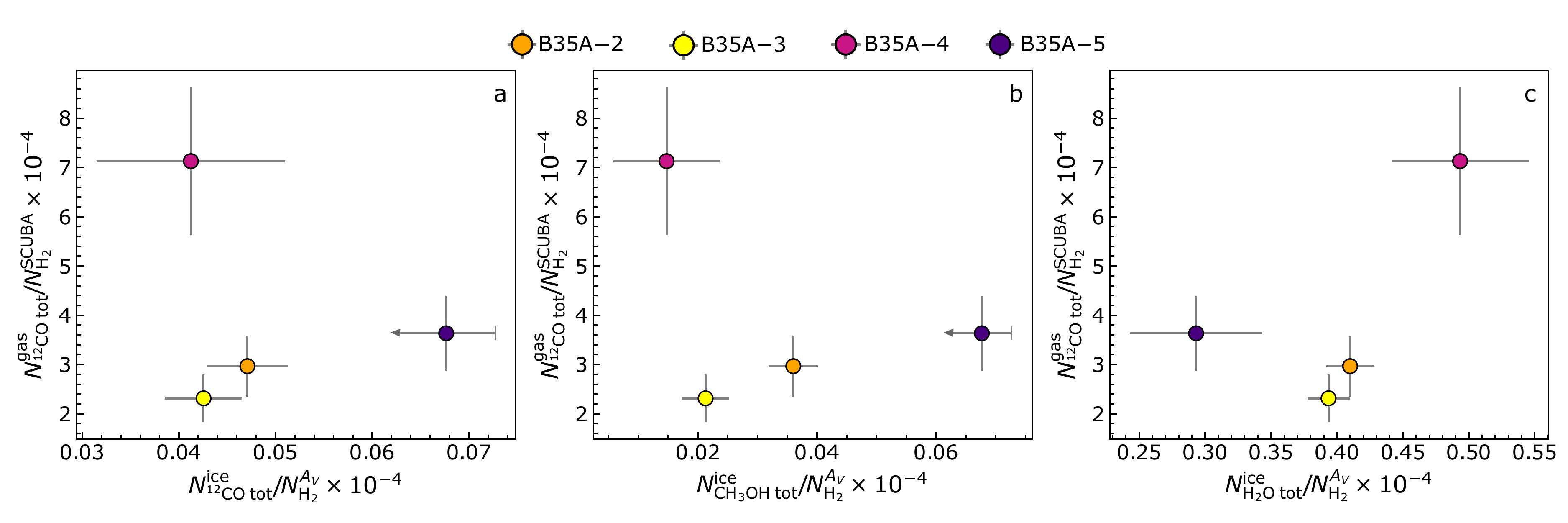}
\caption{Gas and ice variations in B35A. The circles represent the targeted B35A sources. Panels (a)~$-$~(c) show the relation between CO, CH$_3$OH, and H$_2$O ice and CO gas abundances relative to H$_2$.}
\label{Corr_gas_ice}
\end{figure*}

In Figure~\ref{ig_maps}~(c), the distribution of gas-phase CH$_3$OH 5$_{0}-$4$_{0}$ A$^+$ emission is compared to CH$_3$OH ice abundances. As described in Section~\ref{Gas_cd}, the CH$_3$OH emission is concentrated solely in one ridge, located in proximity of the B35A$-$5 position. At this location only an upper limit on the CH$_3$OH ice abundance could be determined. The morphology of the CH$_3$OH emission suggests that while B35A$-$5 is situated in a shocked region, sitting along the trajectory of the giant outflow, B35A$-$4 lies in a more quiescent area, less influenced by high-velocity flows of matter.
The CH$_3$OH ice abundances with respect to H$_2$ are characterized by slightly larger variations compared to CO ice (Table~\ref{table:summary_abund}).

Finally, in Figure~\ref{ig_maps}~(d), the distribution of gas-phase CH$_3$OH 5$_{0}-$4$_{0}$ A$^{+}$ emission is compared to H$_2$O ice abundances. The H$_2$O ice abundances are approximately one order of magnitude higher than the CO and CH$_3$OH ice abundances (Table~\ref{table:summary_abund}). The lowest H$_2$O ice abundance is reported towards B35A$-$5 where 
the CH$_3$OH peak emission is observed. In contrast, the highest H$_2$O ice abundance is obtained towards B35A$-$4, where CH$_3$OH emission is not detected. 


\subsection{Gas and ice variations}
\label{Gas_and_ice_variations}

In Figure~\ref{Corr_gas_ice}, the search for gas-ice correlations towards B35A is addressed by analysing gas and ice abundances relative to H$_2$ (Table~\ref{table:summary_abund}). The y-axes of all three panels display CO gas abundances, which are compared to abundances of CO ice (panel~a), CH$_3$OH ice (panel~b) and H$_2$O ice (panel~c). To maintain gas and ice observations in their own reference frame, the CO gas abundances have been estimated using $N_\mathrm{H_2}$ calculated from SCUBA$-$2 850~$\mu$m maps, whereas the ice abundances have been calculated using $N_\mathrm{H_2}$ obtained from the visual extinction map (see Section~\ref{H2_cd} and Appendix~\ref{appendixB}). The formalism adopted to propagate the uncertainty from the column densities to the abundances is reported in Appendix~\ref{gas-phase_cd_formalism}.

It can be immediately noted that none of the ice species exhibits a predictable trend in ice abundance with gas abundance. Panels~(a) and (b) show a similar behaviour since they are comparing molecules chemically linked - CO gas versus CO ice and CO gas vs CH$_3$OH ice being CO the precursor of CH$_3$OH. At the same time, panel (c) does not display the same relationships seen in the previous two panels. The latter observation is not surprising since there are no reasons for CO and H$_2$O to be directly linked from a purely chemical perspective. 

In both panels~(a) and (b) of Figure~\ref{Corr_gas_ice}, B35A$-$4 displays the highest CO gas and the lowest CO and CH$_3$OH ice abundances. When the uncertainties are taken into account, the CO ice abundances towards B35A$-$2, B35A$-$3 and B35A$-$4 are alike (Table~\ref{table:summary_abund}). For the CH$_3$OH ice abundances, B35A$-$3 and B35A$-$4 are consistent within the error bars, only towards B35A$-$2 a significant difference is found. Both CO and CH$_3$OH ice abundances towards B35A$-$5 are upper limits (Table~\ref{table:summary_abund}). The similarity between CO and CH$_3$OH ice trends and the fact that the CH$_3$OH/CO ice ratios are~$>$~0.3 indicates an efficient CH$_3$OH formation through CO hydrogenation \citep{Watanabe2002} on the grain surfaces of B35A. The CO gas abundances towards B35A$-$2, B35A$-$3 and B35A$-$5 are in agreement within the uncertainties (Table~\ref{table:summary_abund}).

In panels~(c) some of the gas-ice variations differ with respect to panels (a) and (b): B35A$-$4 is characterized by the highest H$_2$O ice abundance, whereas B35A$-$5 shows the lowest H$_2$O ice abundance. This result corroborates the prediction that the ices of B35A$-$5 are the most depleted of volatile molecules among the B35A sources. The behaviours of B35A$-$2 and B35A$-$3 are similar if the uncertainties are considered (Table~\ref{table:summary_abund}). The observed ice and gas variations between CO gas, CH$_3$OH gas and H$_2$O ice towards B35A$-$4 and B35A$-$5 favour a scenario in which H$_2$O ice is formed and predominantly resides on the ices of B35A$-$4, likely located in the most "quiescent" area of the targeted region, whereas in the proximity of a shocked region (B35A$-$5), H$_2$O ice mantles are partially desorbed. This result is discussed further in Sect.~\ref{sputtering}. Although the gas and ice variations analysed in this Section and in Sect.~\ref{comb_maps} are based on a small sample, they still highlight interesting trends in accordance to what we would expect from our knowledge of the physical structure of the targeted region.

\subsection{Comparison with previous gas-ice maps of B35A}

\citet{Noble2017} produced the first gas-ice maps of B35A.
They compared AKARI ice abundances from \citet{Noble2013} with single-dish (JCMT/HARP and IRAM~30~m/HERA) gas-phase maps of C$^{18}$O (\citealt{Buckle2009, Craigon2015}; $J$=3$-$2 and 2$-$1), submillimeter continuum SCUBA maps at 850~$\mu$m, 450~$\mu$m/850~$\mu$m \citep{DiFrancesco2008} and Herschel/SPIRE maps at 250~$\mu$m \citep{Griffin2010}. The H$_2$ gas-phase map adopted to calculate ice abundances was determined from CO observations \citep{Craigon2015}. 

\begin{figure}
\centering
\includegraphics[trim={12 10 5 10},clip,width=3.in]{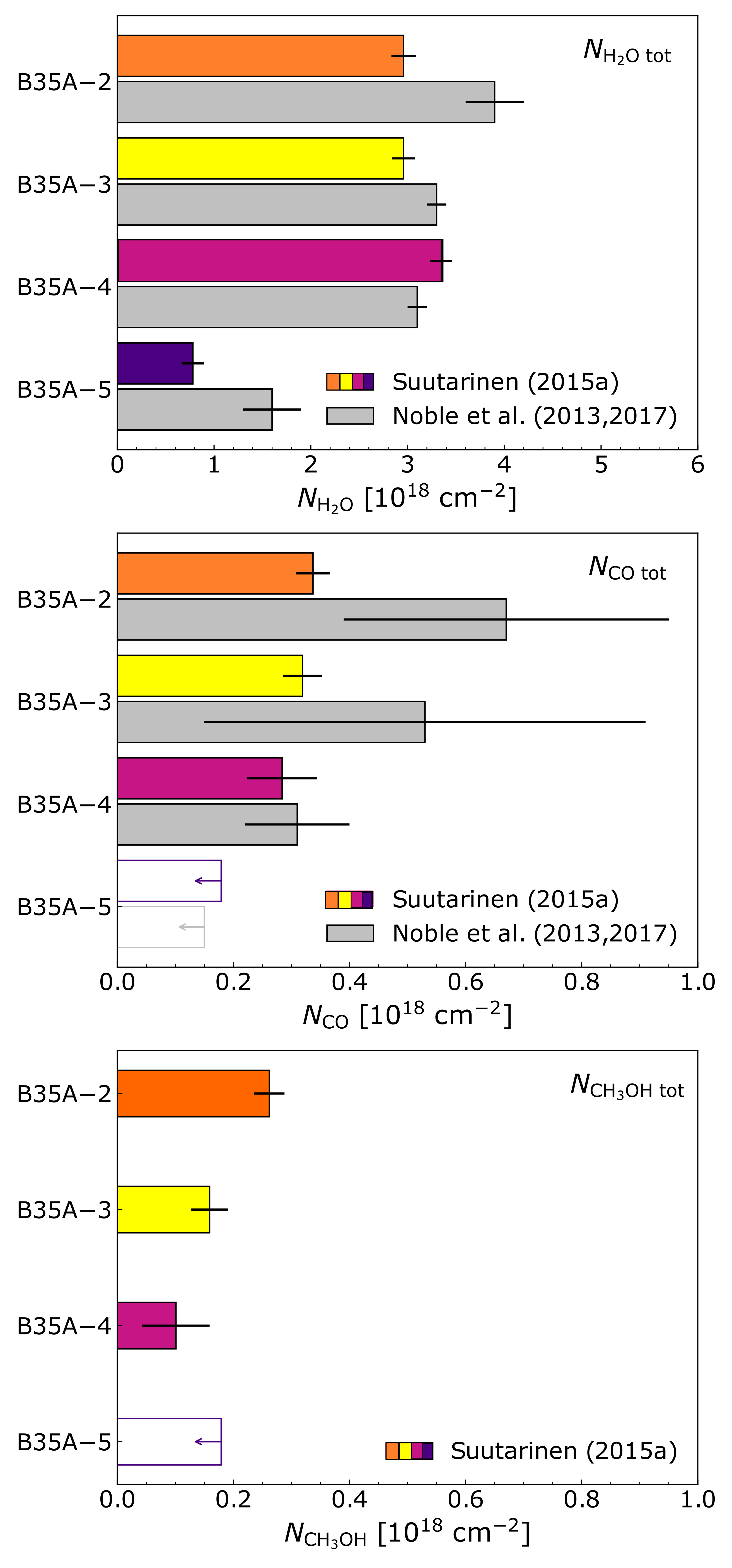}
\caption{H$_2$O (top), CO (middle) and CH$_3$OH (bottom) total ice column densities obtained in \citet{Suutarinen2015} (the sources are colour-coded) compared to \citet{Noble2017} (light grey). The empty bars represent upper limits on the column densities.}
\label{barplot}
\end{figure}

\citet{Noble2013,Noble2017} derived their ice column densities without considering the presence of CH$_3$OH, thus the H$_2$O ice column densities were routinely higher than those used here from \citet{Suutarinen2015} (Figure~\ref{barplot}). Simultaneously, the CO ice column densities were calculated using different laboratory data (e.g., the CO-ice red component was fitted using a CO:H$_2$O \citep{Fraser&Dishoeck2004} laboratory spectrum as opposite to the CO:CH$_3$OH mixture \citep{Cuppen2011}, resulting in higher CO-ice column density compared to \citet{Suutarinen2015}.

The combination of gas, dust and ice observations proposed in \citet{Noble2017} is characterized by a number of limitations which are taken into account in the presented work. The lower angular resolution of the adopted single-dish observations, the analysis of the chemical behaviours of simple molecules only (e.g., H$_2$O, CO and CO$_2$) and the avoidance to consider that the ice and gas observations might probe different depths are some of them. The presented work overcomes these limitations by making use of higher angular resolution interferometric observations, studying the ice and gas variations of both simple (H$_2$O, CO) and complex molecules (CH$_3$OH) and lastly, by comparing ice and gas abundances derived using two different H$_2$ column density maps (to keep the ice and gas observations in their own reference frame). 

From the analysis of the gas-ice maps, \citet{Noble2017} inferred that the dust in B35A is mainly localized around B35A$-$2 and B35A$-$3, in agreement with the higher ice column densities of H$_2$O and CO towards these two sources (Figure~\ref{barplot}). However, they observed that the H$_2$O ice column densities are lower towards B35A$-$5 compared to B35A$-$4, even though the dust emission is higher towards B35A$-$5. The authors concluded that no clear trend is seen between ice, dust and gas in B35A and that more knowledge of the local-scale astrophysical environment is needed to disentangle the exact interplay between ice and gas in the region. 
The findings of \citet{Noble2017} are in agreement with the observational results and the gas-ice maps of B35A presented here (with the caveat that both CO ice abundances towards B35A$-$5 in \citet{Noble2017} and in this work are upper limits). However, the gas-ice maps produced in this study allow to explain that the lower H$_2$O ice abundance towards B35A$-$5 compared to B35A$-$4 is plausibly caused by the influence of shocks on the ice chemistry. CH$_3$OH is a tracer of energetic inputs in the form of outflows \citep{Kristensen2010, Suutarinen2014}, and it traces the trajectory of the bipolar outflow in B35A. As a result, lower H$_2$O ice abundances towards B35A$-$5 are likely the result of sputtering effects in the outflow shocks (e.g., \citealt{Suutarinen2014} and references therein). Simultaneously, the fact that the submillimeter dust emission towards B35A$-$5 is high where the H$_2$O ice abundances are lower can also be explained by the presence of the outflow terminating in the HH~object 175. In fact, UV radiation generated in the Herbig-Haro object itself may locally heat the dust, enhancing the sub-millimeter continuum flux of the region (\citealt{Chandler1996,Kristensen2010}; Figure~\ref{NH2_maps}~a). 


\section{Discussion}
\label{Discussion}
The CH$_3$OH gas abundances in B35A (10$^{-8}$) are greater than what could be inferred by pure gas-phase synthesis (e.g., \citealt{Garrod2006}). Hence, the observed CH$_3$OH gas must be produced on the grain surfaces and be desorbed afterwards. In the following, the preferred scenario for CH$_3$OH desorption in B35A is discussed. Additionally, the observational results presented in Sections~\ref{Gas_cd}$-$\ref{Ice_cd} are employed to determine CH$_3$OH and CO gas-to-ice ratios ($N_\mathrm{gas}$/$N_\mathrm{ice}$) towards the multiple protostellar system in B35A. Finally, a comparison between the gas-to-ice ratios directly determined towards B35A and the gas-to-ice ratios of nearby star-forming regions is proposed.  

\subsection{Sputtering of CH$_3$OH in B35A}
\label{sputtering}

In Sections~\ref{Gas_cd} and~\ref{comb_maps}, it is seen that the CH$_3$OH emission is concentrated almost exclusively in one ridge, which coincides with the B35A$-$5 position. B35A$-$5 sits along the eastern lobe of a large collimated outflow emanated from the binary Class~I IRAS~05417+0907 (i.e, B35A$-3$). The eastern lobe has a total extent of 0.6~pc and it terminates into the Herbig-Haro object HH~175 \citep{Craigon2015}. The total extent of the outflow is 1.65~pc \citep{Reipurth2020}, which places HH~175 among the several dozens HH objects with parsec dimensions \citep{Reipurth2001}. 
Herbig-Haro objects are omnipresent in star-forming regions, but it is still not clear how exactly they are tied into the global star-formation process, e.g., how do they alter the morphology and kinematics of the gas in which they originate. These are highly energetic phenomena which are believed to form when high-velocity protostellar jets collide with the surrounding molecular cloud, inducing shocks. Such shocks compress and heat the gas, generating UV radiation \citep{Neufeld1989}. The activity of Herbig-Haro objects can dramatically change the chemical composition of the gas in their vicinity, mutating the chemistry of the region during so-called sputtering processes \citep{Neufeld1989,Hollenbach1989}.

\begin{figure*}
\centering
\includegraphics[trim={0 0 0 0},clip,width=7.35in]{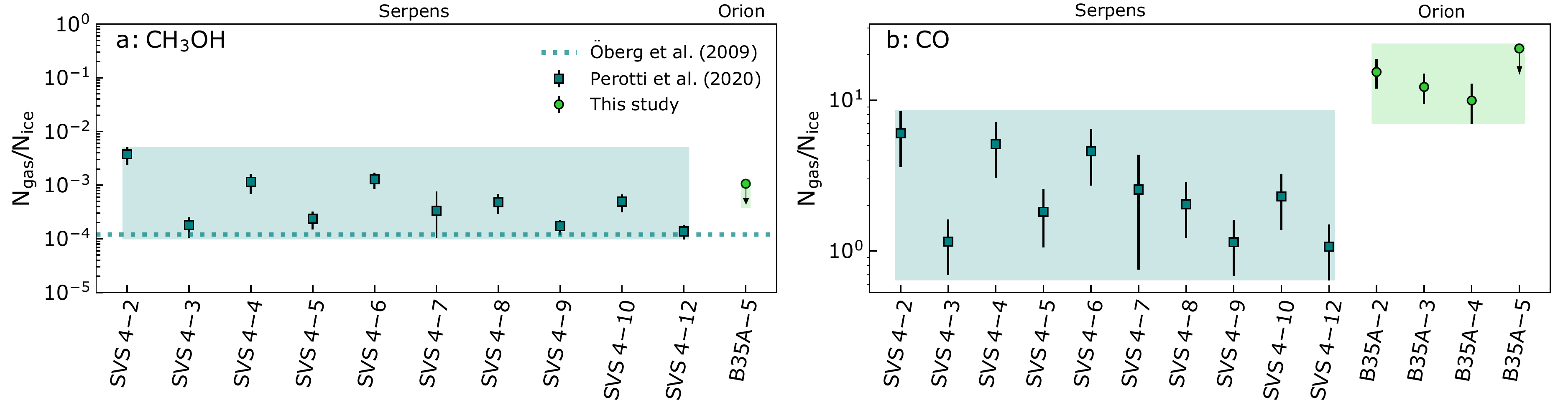}
\caption{CH$_3$OH (panel a) and CO (panel b) gas-to-ice ratios ($N_\mathrm{gas}$/$N_\mathrm{ice}$) for the multiple protostellar system in Orion B35A and Serpens SVS~4. The dark green squares indicate the gas-to-ice ratios estimated by \citet{Perotti2020} for the Serpens SVS~4 cluster. The light green circles mark the ratios calculated in this study. The dotted light blue line in panel~(a) represents the ratio estimated by \citet{Oberg2009a} towards Class 0/I objects. The shaded areas indicate the estimated ranges of the gas-to-ice ratios.}
\label{gas_ice_ratio}
\end{figure*}

Sputtering takes place in shocks when neutral species (e.g., H$_2$, H or He) collide with the surface of ice-covered dust grains with sufficient kinetic energy to expel ice species (e.g., CH$_3$OH, NH$_3$, H$_2$O) into the gas phase \citep{Jorgensen2004, Jimenez-Serra2008, Kristensen2010, Suutarinen2014, Allen2020}. During a sputtering event, the ice molecules can desorb intact or be destroyed (e.g., via dissociative desorption) either due to the high kinetic energy or by reactions with H atoms \citep{Blanksby2003,Dartois2019,Dartois2020}. The fragments of dissociated  molecules can then readily recombine in the gas-phase (\citealt{Suutarinen2014} and references therein).  

The inferred dust temperatures in the B35A region (20$-$30~K) are remarkably lower than the CH$_3$OH sublimation temperature ($\sim$128~K; \citealt{Penteado2017}), excluding thermal desorption as potential mechanism triggering the observed CH$_3$OH emission. This consideration, together with the observation of the CH$_3$OH emission along the HH~175 flow direction and the anticorrelation between CH$_3$OH emission and the lower H$_2$O ice abundances towards B35A$-$5 suggests that ice sputtering in shock waves is a viable mechanism to desorb ice molecules in B35A. Simultaneously, the absence of shocks towards B35A$-$4 likely explains the highest abundance of H$_2$O ice observed in this more quiescent region (Figure~\ref{ig_maps}~d). Higher angular resolution observations of CH$_3$OH in the region are required to provide quantitative measurements of the amount of CH$_3$OH desorbed via sputtering compared to other non-thermal desorption mechanisms. Additional gas-phase observations of H$_2$O towards the B35A sources are needed to constrain further the processes linking H$_2$O ice and H$_2$O gas in this shocked region (e.g., sputtering of H$_2$O ice mantles versus direct H$_2$O gas-phase synthesis) and, in a broader context to contribute explaining the high abundances of H$_2$O observed in molecular outflows \citep{Nisini2010,Bjerkeli2012,Bjerkeli2016,Santangelo2012,Vasta2012, Dionatos2013}.   
\subsection{Gas-to-ice ratios}

Figure~\ref{gas_ice_ratio} displays the CH$_3$OH (panel~a) and CO (panel~b) gas-to-ice ratios towards ten sources constituting the SVS~4 cluster located in the Serpens molecular cloud \citep{Perotti2020} and towards the B35A sources in the Orion molecular cloud. The shaded areas indicate the estimated ranges of gas-to-ice ratios towards both molecular clouds, providing complementary information of the gas-to-ice ratios of both regions compared to the value calculated towards each individual source.   

Figure~\ref{gas_ice_ratio}~(a) shows the CH$_3$OH gas-to-ice ratios ($N_\mathrm{CH_3OH_{gas}}$/$N_\mathrm{CH_3OH_{ice}}$) towards the SVS~4 sources and B35A$-$5. Only an upper limit for the CH$_3$OH gas-to-ice towards B35A$-$5 is obtained (i.e, 1.1~$\times$~10$^{-3}$) due to the uncertainty on the CH$_3$OH ice column density determination towards this source \citep{Suutarinen2015}. Therefore, only limited information is available on the CH$_3$OH chemistry in B35A. Figure~\ref{gas_ice_ratio}~(a) also displays the averaged CH$_3$OH gas-to-ice ratio towards four low-mass embedded protostars located in nearby star-forming regions from \citet{Oberg2009a}. A detailed comparison between the CH$_3$OH gas-to-ice ratios for the SVS~4 cluster members and the value estimated by \citet{Oberg2009a} is given in \citet{Perotti2020}, and therefore not repeated here.

The observations of B35A address cold quiescent CH$_3$OH emission ($J$= 5$_K-$4$_K$ transitions at 241.7~GHz and $E_\mathrm{u}$ from 34.8 to 60.7~K, see Table~\ref{table:spectral_data_pointings}) and presumably quiescent gas not affected by outflow shocks and/or thermally desorbed in the vicinity of the protostar. The same applies to the observations presented in \citet{Oberg2009a} ($J$=~2$-$1 transitions at 96.7~GHz and $E_\mathrm{u}$ in the range 14.4$-$35.4~K). However, B35A$-$5 and the SVS~4 sources reporting the highest gas-to-ice ratios are likely affected to some extent by outflow shocks (i.e., B35A$-$5 by the giant outflow terminating in HH~175 and the SVS~4 sources by the outflow notoriously associated to SMM~4). This might justify the mismatch between \citet{Oberg2009a} value and the higher values of the gas-to-ice ratios. In the case of the B35A cloud, high-sensitivity infrared observations of B35A are required to constrain further this observation.  

CO gas-to-ice ratios ($N_\mathrm{CO_{gas}}$/$N_\mathrm{CO_{ice}}$) towards the SVS~4 and B35A sources are illustrated in Figure~\ref{gas_ice_ratio}~(b). As discussed previously in \citet{Perotti2020}, the range of gas-to-ice ratios for SVS~4 extends between 1 and 6 and it is significantly higher than the predicted value ($\sim$4$\times$10$^{-2}$; \citealt{Cazaux2016}). This theoretical prediction is estimated from the three-phase astrochemical model by \citet{Cazaux2016} which includes thermal and non-thermal desorption processes of the species constituting the ice mantles. This comparison with the theoretical prediction has to be taken with care, since the physical model is not specifically tuned to reproduce the structures of the SVS~4 cluster or the B35A cloud. The estimated dust temperature towards the SVS~4 cluster is below 20~K \citep{Kristensen2010}, indicating that CO is likely frozen out on the grain surfaces in this temperature range. We concluded from the high relative CO gas abundances that the gas-mapping is not tracing the densest regions of the SVS~4 cluster but an extended component that is not sensitive to the effect of freeze-out (Figure~8 of \citealt{Perotti2020}). 

High CO gas-to-ice ratios are also observed for B35A, but in this case they are mainly attributable to CO thermally desorbed in the region. The physical conditions of B35A and SVS~4 differ significantly, for example, the western side of B35A is highly influenced by the ionization-shock front driven by the $\lambda$~Orionis OB stars and the dust temperature in the YSOs region of B35A has been estimated to be at least 25$-$30~K \citep{Craigon2015, Morgan2008}. Therefore, at these temperatures, above the CO sublimation temperature ($\sim$20~K; \citealt{Bisschop2006,Collings2004,Acharyya2007}), CO is efficiently thermally desorbed from the grain surfaces and it is expected to primarily reside in the gas-phase. According to this result, the observed CH$_3$OH ice might have formed at earlier stages, prior to the warm-up of the dust grains or alternatively, from CO molecules trapped in porous H$_2$O matrices, and therefore not sublimated at these temperatures. Additionally, higher abundances of CO might reflect an active CO gas-phase synthesis in the region. Follow-up observations of higher $J$ transitions of CO isotopologues are required to provide a conclusive assessment.


\section{Conclusions}
\label{Conclusions}

Millimetric (SMA, APEX, IRAM~30~m) and infrared (AKARI) observations are used to investigate the relationships between ice, dust and gas in the B35A cloud, associated with the $\lambda$~Orionis region. Gas and ice maps are produced to compare the distribution of solid (H$_2$O, CO and CH$_3$OH) and gaseous ($^{13}$CO, C$^{18}$O and CH$_3$OH) molecules and to link the small-scale variations traced by ice observations with larger-scale astrophysical phenomena probed by gas observations. The main conclusions are:
\begin{itemize}
\item{The CO isotopologues emission is extended in B35A, whereas 
the observed CH$_3$OH emission is compact and centered in the vicinity of B35A$-$5. B35A$-$5 sits along the trajectory of the outflow emanated from IRAS~05417+0907 (i.e., B35A$-$3), thus the observed gas-phase CH$_3$OH may be plausibly explained by sputtering of ice CH$_3$OH in the outflow shocks.}
\item{The dust column density traced by the submillimeter emission is not directly related to the ice column inferred from the infrared observations. The submillimeter dust emission is stronger towards B35A$-$2, B35A$-$3 and B35A$-$5 compared to B35A$-$4, however the H$_2$O ice column density is higher for B35A$-$4. This discrepancy is understood by taking into account that B35A$-$2, B35A$-$3 and B35A$-$5 are situated in a shocked region affected by the presence of the Herbig-Haro object HH175 -- and thus likely influenced by sputtering and heating affecting the observed submillimeter dust emission pattern.}
\item{None of the ice species shows a predictable trend in ice abundance with gas-phase abundance. This implies that inferring ice abundances from known gas-phase abundances and vice versa is inaccurate without an extensive knowledge of the physical environment of the targeted region.}
\item{The high CO gas-to-ice ratios suggest that the CO molecules are efficiently thermally desorbed in B35A. This is supported by dust temperature 
estimates (25$-$30~K) towards the B35A sources above the CO sublimation temperature ($\sim$20~K).}
\item{Simultaneously, the dust temperatures in the region are significantly lower than the CH$_3$OH sublimation temperature (128~K), excluding thermal desorption as the mechanism responsible for the observed CH$_3$OH emission.}
\item{The combination of gas- and ice observations is essential to comprehend the relationships at the interface between solid and gas phases, and hence to link the small-scale variations detected in the ice observations with large-scale phenomena revealed by gas-phase observations.}
\end{itemize}
The presented work is a preparatory study for future JWST~+~ALMA observations, which will shed further light on the dependencies of gas-to-ice ratios on the physical conditions of star-forming regions. In fact, future mid-infrared facilities, among all JWST, will considerably increase the number of regions for which ice maps are available. Such high-sensitivity ice maps can be then combined with ALMA observations, to provide better constraints on the complex interplay between ice, dust and gas during the earliest phases of star-formation. The present study already shows that such gas-ice maps will be valuable, given that they probe ice- and gas-phase chemistries to a greater extent than by ice- or gas-maps alone. 

\begin{acknowledgements}
The authors wish to thank Bo Reipurth for fruitful discussions on HH~175 and Alison Craigon, Zak Smith, Jennifer Noble for providing the reduced IRAM~30~m data. The authors also wish to acknowledge the significance of Mauna Kea to the indigenous Hawaiian people and the anonymous reviewer for the careful reading of the manuscript and the useful comments. This work is based on observations with the Submillimeter Array, Mauna Kea, Hawaii, program code: 2018A-S033, with the Atacama Pathfinder Experiment, Llano Chajnantor, Chile, program code: 0102.F-9304. The Submillimeter Array is a joint project between the Smithsonian Astrophysical Observatory and the Academia Sinica Institute of Astronomy and Astrophysics and is funded by the Smithsonian Institution and the Academia Sinica. The Atacama Pathﬁnder EXperiment (APEX) telescope is a collaboration between the Max Planck Institute for Radio Astronomy, the European Southern Observatory, and the Onsala Space Observatory. Swedish observations on APEX are supported through Swedish Research Council grant No 2017-00648. The study is also based on data from the IRAM Science Data Archive, obtained by H.J.Fraser with the IRAM~30~m telescope under project ID~088-07. Finally, this work is based on archival data from the AKARI satellite, a JAXA project with the participation of the European Space Agency (ESA). The group of JKJ acknowledges the financial support from the European Research Council (ERC) under the European Union's Horizon 2020 research and innovation programme (grant agreement No 646908) through ERC Consolidator Grant "S4F". The research of LEK is supported by research grant (19127) from VILLUM FONDEN. HJF gratefully acknowledges the support of STFC for Astrochemistry at the OU under grant Nos ST/P000584/1 and ST/T005424/1 enabling her participation in this work.
\end{acknowledgements}


\bibliographystyle{aa}
\bibliography{References}

\begin{thebibliography}{118}
\expandafter\ifx\csname natexlab\endcsname\relax\def\natexlab#1{#1}\fi

\bibitem[{{Acharyya} {et~al.}(2007){Acharyya}, {Fuchs}, {Fraser}, {van
  Dishoeck}, \& {Linnartz}}]{Acharyya2007}
{Acharyya}, K., {Fuchs}, G.~W., {Fraser}, H.~J., {van Dishoeck}, E.~F., \&
  {Linnartz}, H. 2007, \aap, 466, 1005

\bibitem[{{Allen} {et~al.}(2020){Allen}, {Cordiner}, \& {Charnley}}]{Allen2020}
{Allen}, V., {Cordiner}, M., \& {Charnley}, S. 2020, arXiv e-prints,
  arXiv:2010.01151

\bibitem[{{Ansdell} {et~al.}(2020){Ansdell}, {Haworth}, {Williams}, {Facchini},
  {Winter}, {Manara}, {Hacar}, {Chiang}, {van Terwisga}, {van der Marel}, \&
  {van Dishoeck}}]{Ansdell2020}
{Ansdell}, M., {Haworth}, T.~J., {Williams}, J.~P., {et~al.} 2020, \aj, 160,
  248

\bibitem[{{Barnes} {et~al.}(2015){Barnes}, {Muller}, {Indermuehle},
  {O'Dougherty}, {Lowe}, {Cunningham}, {Hernandez}, \& {Fuller}}]{Barnes2015}
{Barnes}, P.~J., {Muller}, E., {Indermuehle}, B., {et~al.} 2015, \apj, 812, 6

\bibitem[{{Barrado} {et~al.}(2018){Barrado}, {de Gregorio Monsalvo},
  {Hu{\'e}lamo}, {Morales-Calder{\'o}n}, {Bayo}, {Palau}, {Ruiz},
  {Rivi{\`e}re-Marichalar}, {Bouy}, {Morata}, {Stauffer}, {Eiroa}, \&
  {Noriega-Crespo}}]{Barrado2018}
{Barrado}, D., {de Gregorio Monsalvo}, I., {Hu{\'e}lamo}, N., {et~al.} 2018,
  \aap, 612, A79

\bibitem[{{Barrado} {et~al.}(2011){Barrado}, {Stelzer}, {Morales-Calder{\'o}n},
  {Bayo}, {Hu{\'e}lamo}, {Stauffer}, {Hodgkin}, {Galindo}, \&
  {Verdugo}}]{Barrado2011}
{Barrado}, D., {Stelzer}, B., {Morales-Calder{\'o}n}, M., {et~al.} 2011, \aap,
  526, A21

\bibitem[{{Bayo} {et~al.}(2011){Bayo}, {Barrado}, {Stauffer},
  {Morales-Calder{\'o}n}, {Melo}, {Hu{\'e}lamo}, {Bouy}, {Stelzer}, {Tamura},
  \& {Jayawardhana}}]{Bayo2011}
{Bayo}, A., {Barrado}, D., {Stauffer}, J., {et~al.} 2011, \aap, 536, A63

\bibitem[{Bergin \& Tafalla(2007)}]{Bergin2007}
Bergin, E.~A. \& Tafalla, M. 2007, Annual Review of Astronomy and Astrophysics,
  45, 339–396

\bibitem[{{Bergner} {et~al.}(2017){Bergner}, {{\"O}berg}, {Garrod}, \&
  {Graninger}}]{Bergner2017}
{Bergner}, J.~B., {{\"O}berg}, K.~I., {Garrod}, R.~T., \& {Graninger}, D.~M.
  2017, \apj, 841, 120

\bibitem[{{Bertin} {et~al.}(2016){Bertin}, {Romanzin}, {Doronin}, {Philippe},
  {Jeseck}, {Ligterink}, {Linnartz}, {Michaut}, \& {Fillion}}]{Bertin2016}
{Bertin}, M., {Romanzin}, C., {Doronin}, M., {et~al.} 2016, \apjl, 817, L12

\bibitem[{{Bisschop} {et~al.}(2006){Bisschop}, {Fraser}, {{\"O}berg}, {van
  Dishoeck}, \& {Schlemmer}}]{Bisschop2006}
{Bisschop}, S.~E., {Fraser}, H.~J., {{\"O}berg}, K.~I., {van Dishoeck}, E.~F.,
  \& {Schlemmer}, S. 2006, \aap, 449, 1297

\bibitem[{{Bjerkeli} {et~al.}(2016){Bjerkeli}, {J{\o}rgensen}, {Bergin},
  {Frimann}, {Harsono}, {Jacobsen}, {Lindberg}, {Persson}, {Sakai}, {van
  Dishoeck}, {Visser}, \& {Yamamoto}}]{Bjerkeli2016}
{Bjerkeli}, P., {J{\o}rgensen}, J.~K., {Bergin}, E.~A., {et~al.} 2016, \aap,
  595, A39

\bibitem[{{Bjerkeli} {et~al.}(2012){Bjerkeli}, {Liseau}, {Larsson}, {Rydbeck},
  {Nisini}, {Tafalla}, {Antoniucci}, {Benedettini}, {Bergman}, {Cabrit},
  {Giannini}, {Melnick}, {Neufeld}, {Santangelo}, \& {van
  Dishoeck}}]{Bjerkeli2012}
{Bjerkeli}, P., {Liseau}, R., {Larsson}, B., {et~al.} 2012, \aap, 546, A29

\bibitem[{Blanksby \& Ellison(2003)}]{Blanksby2003}
Blanksby, S. \& Ellison, G. 2003, Accounts of Chemical Research, 36, 255

\bibitem[{{Boogert} {et~al.}(2015){Boogert}, {Gerakines}, \&
  {Whittet}}]{Boogert2015}
{Boogert}, A.~C.~A., {Gerakines}, P.~A., \& {Whittet}, D.~C.~B. 2015, \araa,
  53, 541

\bibitem[{{Buckle} {et~al.}(2009){Buckle}, {Hills}, {Smith}, {Dent}, {Bell},
  {Curtis}, {Dace}, {Gibson}, {Graves}, {Leech}, {Richer}, {Williamson},
  {Withington}, {Yassin}, {Bennett}, {Hastings}, {Laidlaw}, {Lightfoot},
  {Burgess}, {Dewdney}, {Hovey}, {Willis}, {Redman}, {Wooff}, {Berry},
  {Cavanagh}, {Davis}, {Dempsey}, {Friberg}, {Jenness}, {Kackley}, {Rees},
  {Tilanus}, {Walther}, {Zwart}, {Klapwijk}, {Kroug}, \&
  {Zijlstra}}]{Buckle2009}
{Buckle}, J.~V., {Hills}, R.~E., {Smith}, H., {et~al.} 2009, \mnras, 399, 1026

\bibitem[{{Calcutt} {et~al.}(2018){Calcutt}, {J{\o}rgensen}, {M{\"u}ller},
  {Kristensen}, {Coutens}, {Bourke}, {Garrod}, {Persson}, {van der Wiel}, {van
  Dishoeck}, \& {Wampfler}}]{Calcutt2018}
{Calcutt}, H., {J{\o}rgensen}, J.~K., {M{\"u}ller}, H.~S.~P., {et~al.} 2018,
  \aap, 616, A90

\bibitem[{{Carlhoff} {et~al.}(2013){Carlhoff}, {Nguyen Luong}, {Schilke},
  {Motte}, {Schneider}, {Beuther}, {Bontemps}, {Heitsch}, {Hill}, {Kramer},
  {Ossenkopf}, {Schuller}, {Simon}, \& {Wyrowski}}]{Carlhoff2013}
{Carlhoff}, P., {Nguyen Luong}, Q., {Schilke}, P., {et~al.} 2013, \aap, 560,
  A24

\bibitem[{{Cazaux} {et~al.}(2017){Cazaux}, {Mart{\'\i}n-Dom{\'e}nech}, {Chen},
  {Mu{\~n}oz Caro}, \& {Gonz{\'a}lez D{\'\i}az}}]{Cazaux2017}
{Cazaux}, S., {Mart{\'\i}n-Dom{\'e}nech}, R., {Chen}, Y.~J., {Mu{\~n}oz Caro},
  G.~M., \& {Gonz{\'a}lez D{\'\i}az}, C. 2017, \apj, 849, 80

\bibitem[{{Cazaux} {et~al.}(2016){Cazaux}, {Minissale}, {Dulieu}, \&
  {Hocuk}}]{Cazaux2016}
{Cazaux}, S., {Minissale}, M., {Dulieu}, F., \& {Hocuk}, S. 2016, \aap, 585,
  A55

\bibitem[{{Cazaux} {et~al.}(2003){Cazaux}, {Tielens}, {Ceccarelli}, {Castets},
  {Wakelam}, {Caux}, {Parise}, \& {Teyssier}}]{Cazaux2003}
{Cazaux}, S., {Tielens}, A.~G.~G.~M., {Ceccarelli}, C., {et~al.} 2003, \apjl,
  593, L51

\bibitem[{{Chandler} \& {Carlstrom}(1996)}]{Chandler1996}
{Chandler}, C.~J. \& {Carlstrom}, J.~E. 1996, \apj, 466, 338

\bibitem[{{Chapman} {et~al.}(2009){Chapman}, {Mundy}, {Lai}, \&
  {Evans}}]{Chapman2009}
{Chapman}, N.~L., {Mundy}, L.~G., {Lai}, S.-P., \& {Evans}, II, N.~J. 2009,
  \apj, 690, 496

\bibitem[{{Collings} {et~al.}(2004){Collings}, {Anderson}, {Chen}, {Dever},
  {Viti}, {Williams}, \& {McCoustra}}]{Collings2004}
{Collings}, M.~P., {Anderson}, M.~A., {Chen}, R., {et~al.} 2004, \mnras, 354,
  1133

\bibitem[{{Connelley} {et~al.}(2008){Connelley}, {Reipurth}, \&
  {Tokunaga}}]{Connelley2008}
{Connelley}, M.~S., {Reipurth}, B., \& {Tokunaga}, A.~T. 2008, \aj, 135, 2496

\bibitem[{{Conti} \& {Leep}(1974)}]{Conti1974}
{Conti}, P.~S. \& {Leep}, E.~M. 1974, \apj, 193, 113

\bibitem[{Craigon(2015)}]{Craigon2015}
Craigon, A.~M. 2015, PhD thesis, Dept. of Physics, Univ. of Strathclyde,
  \url{http://digitool.lib.strath.ac.uk/R/?func=dbin-jump-full&object_id=27550}

\bibitem[{{Cruz-Diaz} {et~al.}(2016){Cruz-Diaz}, {Mart{\'{\i}}n-Dom{\'e}nech},
  {Mu{\~n}oz Caro}, \& {Chen}}]{Cruz-Diaz2016}
{Cruz-Diaz}, G.~A., {Mart{\'{\i}}n-Dom{\'e}nech}, R., {Mu{\~n}oz Caro}, G.~M.,
  \& {Chen}, Y.-J. 2016, \aap, 592, A68

\bibitem[{{Cuppen} {et~al.}(2011){Cuppen}, {Penteado}, {Isokoski}, {van der
  Marel}, \& {Linnartz}}]{Cuppen2011}
{Cuppen}, H.~M., {Penteado}, E.~M., {Isokoski}, K., {van der Marel}, N., \&
  {Linnartz}, H. 2011, \mnras, 417, 2809

\bibitem[{{Dartois} {et~al.}(2020){Dartois}, {Chabot}, {Bacmann}, {Boduch},
  {Domaracka}, \& {Rothard}}]{Dartois2020}
{Dartois}, E., {Chabot}, M., {Bacmann}, A., {et~al.} 2020, \aap, 634, A103

\bibitem[{{Dartois} {et~al.}(2019){Dartois}, {Chabot}, {Id Barkach}, {Rothard},
  {Aug{\'e}}, {Agnihotri}, {Domaracka}, \& {Boduch}}]{Dartois2019}
{Dartois}, E., {Chabot}, M., {Id Barkach}, T., {et~al.} 2019, Astronomy and
  Astrophysics, 627, A55

\bibitem[{{De Vries} {et~al.}(2002){De Vries}, {Narayanan}, \&
  {Snell}}]{DeVries2002}
{De Vries}, C.~H., {Narayanan}, G., \& {Snell}, R.~L. 2002, \apj, 577, 798

\bibitem[{{Dempsey} {et~al.}(2013){Dempsey}, {Friberg}, {Jenness}, {Tilanus},
  {Thomas}, {Holland}, {Bintley}, {Berry}, {Chapin}, {Chrysostomou}, {Davis},
  {Gibb}, {Parsons}, \& {Robson}}]{Dempsey2013}
{Dempsey}, J.~T., {Friberg}, P., {Jenness}, T., {et~al.} 2013, \mnras, 430,
  2534

\bibitem[{{Di Francesco} {et~al.}(2008){Di Francesco}, {Johnstone}, {Kirk},
  {MacKenzie}, \& {Ledwosinska}}]{DiFrancesco2008}
{Di Francesco}, J., {Johnstone}, D., {Kirk}, H., {MacKenzie}, T., \&
  {Ledwosinska}, E. 2008, \apjs, 175, 277

\bibitem[{{Dionatos} {et~al.}(2013){Dionatos}, {J{\o}rgensen}, {Green},
  {Herczeg}, {Evans}, {Kristensen}, {Lindberg}, \& {van
  Dishoeck}}]{Dionatos2013}
{Dionatos}, O., {J{\o}rgensen}, J.~K., {Green}, J.~D., {et~al.} 2013, \aap,
  558, A88

\bibitem[{{Dolan} \& {Mathieu}(1999)}]{Dolan1999}
{Dolan}, C.~J. \& {Mathieu}, R.~D. 1999, \aj, 118, 2409

\bibitem[{{Dolan} \& {Mathieu}(2002)}]{Dolan2002}
{Dolan}, C.~J. \& {Mathieu}, R.~D. 2002, \aj, 123, 387

\bibitem[{{Dulieu} {et~al.}(2013){Dulieu}, {Congiu}, {Noble}, {Baouche},
  {Chaabouni}, {Moudens}, {Minissale}, \& {Cazaux}}]{Dulieu2013}
{Dulieu}, F., {Congiu}, E., {Noble}, J., {et~al.} 2013, Scientific Reports, 3,
  1338

\bibitem[{{Eistrup} {et~al.}(2016){Eistrup}, {Walsh}, \& {van
  Dishoeck}}]{Eistrup2016}
{Eistrup}, C., {Walsh}, C., \& {van Dishoeck}, E.~F. 2016, \aap, 595, A83

\bibitem[{{Evans} {et~al.}(2014){Evans}, {Allen}, {Blake}, {Boogert}, {Bourke},
  {Harvey}, {Kessler}, {Koerner}, {Lee}, {Mundy}, {Myers}, {Padgett},
  {Pontoppidan}, {Sargent}, {Stapelfeldt}, {van Dishoeck}, {Young}, \&
  {Young}}]{Evans2014}
{Evans}, II, N.~J., {Allen}, L.~E., {Blake}, G.~A., {et~al.} 2014, VizieR
  Online Data Catalog, 2332

\bibitem[{{Evans} {et~al.}(2009){Evans}, {Dunham}, {J{\o}rgensen}, {Enoch},
  {Mer{\'\i}n}, {van Dishoeck}, {Alcal{\'a}}, {Myers}, {Stapelfeldt}, {Huard},
  {Allen}, {Harvey}, {van Kempen}, {Blake}, {Koerner}, {Mundy}, {Padgett}, \&
  {Sargent}}]{Evans2009}
{Evans}, II, N.~J., {Dunham}, M.~M., {J{\o}rgensen}, J.~K., {et~al.} 2009,
  \apjs, 181, 321

\bibitem[{{Fraser} {et~al.}(2001){Fraser}, {Collings}, {McCoustra}, \&
  {Williams}}]{Fraser2001}
{Fraser}, H.~J., {Collings}, M.~P., {McCoustra}, M. R.~S., \& {Williams}, D.~A.
  2001, \mnras, 327, 1165

\bibitem[{{Fraser} \& {van Dishoeck}(2004)}]{Fraser&Dishoeck2004}
{Fraser}, H.~J. \& {van Dishoeck}, E.~F. 2004, Advances in Space Research, 33,
  14

\bibitem[{{Garrod} {et~al.}(2006){Garrod}, {Park}, {Caselli}, \&
  {Herbst}}]{Garrod2006}
{Garrod}, R., {Park}, I.~H., {Caselli}, P., \& {Herbst}, E. 2006, Faraday
  Discussions, 133, 51

\bibitem[{{Goldsmith} \& {Langer}(1999)}]{Goldsmith1999}
{Goldsmith}, P.~F. \& {Langer}, W.~D. 1999, \apj, 517, 209

\bibitem[{{Grassi} {et~al.}(2020){Grassi}, {Bovino}, {Caselli}, {Bovolenta},
  {Vogt-Geisse}, \& {Ercolano}}]{Grassi2020}
{Grassi}, T., {Bovino}, S., {Caselli}, P., {et~al.} 2020, \aap, 643, A155

\bibitem[{{Griffin} {et~al.}(2010){Griffin}, {Abergel}, {Abreu}, {Ade},
  {Andr{\'e}}, {Augueres}, {Babbedge}, {Bae}, {Baillie}, {Baluteau}, {Barlow},
  {Bendo}, {Benielli}, {Bock}, {Bonhomme}, {Brisbin}, {Brockley-Blatt},
  {Caldwell}, {Cara}, {Castro-Rodriguez}, {Cerulli}, {Chanial}, {Chen},
  {Clark}, {Clements}, {Clerc}, {Coker}, {Communal}, {Conversi}, {Cox},
  {Crumb}, {Cunningham}, {Daly}, {Davis}, {de Antoni}, {Delderfield}, {Devin},
  {di Giorgio}, {Didschuns}, {Dohlen}, {Donati}, {Dowell}, {Dowell}, {Duband},
  {Dumaye}, {Emery}, {Ferlet}, {Ferrand}, {Fontignie}, {Fox}, {Franceschini},
  {Frerking}, {Fulton}, {Garcia}, {Gastaud}, {Gear}, {Glenn}, {Goizel},
  {Griffin}, {Grundy}, {Guest}, {Guillemet}, {Hargrave}, {Harwit}, {Hastings},
  {Hatziminaoglou}, {Herman}, {Hinde}, {Hristov}, {Huang}, {Imhof}, {Isaak},
  {Israelsson}, {Ivison}, {Jennings}, {Kiernan}, {King}, {Lange}, {Latter},
  {Laurent}, {Laurent}, {Leeks}, {Lellouch}, {Levenson}, {Li}, {Li},
  {Lilienthal}, {Lim}, {Liu}, {Lu}, {Madden}, {Mainetti}, {Marliani}, {McKay},
  {Mercier}, {Molinari}, {Morris}, {Moseley}, {Mulder}, {Mur}, {Naylor},
  {Nguyen}, {O'Halloran}, {Oliver}, {Olofsson}, {Olofsson}, {Orfei}, {Page},
  {Pain}, {Panuzzo}, {Papageorgiou}, {Parks}, {Parr-Burman}, {Pearce},
  {Pearson}, {P{\'e}rez-Fournon}, {Pinsard}, {Pisano}, {Podosek}, {Pohlen},
  {Polehampton}, {Pouliquen}, {Rigopoulou}, {Rizzo}, {Roseboom}, {Roussel},
  {Rowan-Robinson}, {Rownd}, {Saraceno}, {Sauvage}, {Savage}, {Savini},
  {Sawyer}, {Scharmberg}, {Schmitt}, {Schneider}, {Schulz}, {Schwartz},
  {Shafer}, {Shupe}, {Sibthorpe}, {Sidher}, {Smith}, {Smith}, {Smith},
  {Spencer}, {Stobie}, {Sudiwala}, {Sukhatme}, {Surace}, {Stevens}, {Swinyard},
  {Trichas}, {Tourette}, {Triou}, {Tseng}, {Tucker}, {Turner}, {Vaccari},
  {Valtchanov}, {Vigroux}, {Virique}, {Voellmer}, {Walker}, {Ward}, {Waskett},
  {Weilert}, {Wesson}, {White}, {Whitehouse}, {Wilson}, {Winter}, {Woodcraft},
  {Wright}, {Xu}, {Zavagno}, {Zemcov}, {Zhang}, \& {Zonca}}]{Griffin2010}
{Griffin}, M.~J., {Abergel}, A., {Abreu}, A., {et~al.} 2010, \aap, 518, L3

\bibitem[{{G{\"u}sten} {et~al.}(2006){G{\"u}sten}, {Nyman}, {Schilke},
  {Menten}, {Cesarsky}, \& {Booth}}]{Gusten2006}
{G{\"u}sten}, R., {Nyman}, L.~{\AA}., {Schilke}, P., {et~al.} 2006, \aap, 454,
  L13

\bibitem[{{Heiles} \& {Habing}(1974)}]{Heiles1974}
{Heiles}, C. \& {Habing}, H.~J. 1974, \aaps, 14, 1

\bibitem[{{Herbst} \& {van Dishoeck}(2009)}]{Herbst2009}
{Herbst}, E. \& {van Dishoeck}, E.~F. 2009, \araa, 47, 427

\bibitem[{{Hern{\'a}ndez} {et~al.}(2009){Hern{\'a}ndez}, {Calvet}, {Hartmann},
  {Muzerolle}, {Gutermuth}, \& {Stauffer}}]{Hernandez2009}
{Hern{\'a}ndez}, J., {Calvet}, N., {Hartmann}, L., {et~al.} 2009, \apj, 707,
  705

\bibitem[{{Ho} {et~al.}(2004){Ho}, {Moran}, \& {Lo}}]{Ho2004}
{Ho}, P. T.~P., {Moran}, J.~M., \& {Lo}, K.~Y. 2004, \apjl, 616, L1

\bibitem[{{Hollenbach} \& {McKee}(1989)}]{Hollenbach1989}
{Hollenbach}, D. \& {McKee}, C.~F. 1989, \apj, 342, 306

\bibitem[{{Jim{\'e}nez-Serra} {et~al.}(2008){Jim{\'e}nez-Serra}, {Caselli},
  {Mart{\'\i}n-Pintado}, \& {Hartquist}}]{Jimenez-Serra2008}
{Jim{\'e}nez-Serra}, I., {Caselli}, P., {Mart{\'\i}n-Pintado}, J., \&
  {Hartquist}, T.~W. 2008, \aap, 482, 549

\bibitem[{{J{\o}rgensen} {et~al.}(2020){J{\o}rgensen}, {Belloche}, \&
  {Garrod}}]{Jorgensen2020}
{J{\o}rgensen}, J.~K., {Belloche}, A., \& {Garrod}, R.~T. 2020, \araa, 58, 727

\bibitem[{{J{\o}rgensen} {et~al.}(2004){J{\o}rgensen}, {Hogerheijde}, {Blake},
  {van Dishoeck}, {Mundy}, \& {Sch{\"o}ier}}]{Jorgensen2004}
{J{\o}rgensen}, J.~K., {Hogerheijde}, M.~R., {Blake}, G.~A., {et~al.} 2004,
  \aap, 415, 1021

\bibitem[{{J{\o}rgensen} {et~al.}(2016){J{\o}rgensen}, {van der Wiel},
  {Coutens}, {Lykke}, {M{\"u}ller}, {van Dishoeck}, {Calcutt}, {Bjerkeli},
  {Bourke}, {Drozdovskaya}, {Favre}, {Fayolle}, {Garrod}, {Jacobsen},
  {{\"O}berg}, {Persson}, \& {Wampfler}}]{Jorgensen2016}
{J{\o}rgensen}, J.~K., {van der Wiel}, M.~H.~D., {Coutens}, A., {et~al.} 2016,
  \aap, 595, A117

\bibitem[{{Kauffmann} {et~al.}(2008){Kauffmann}, {Bertoldi}, {Bourke}, {Evans},
  \& {Lee}}]{Kauffmann2008}
{Kauffmann}, J., {Bertoldi}, F., {Bourke}, T.~L., {Evans}, II, N.~J., \& {Lee},
  C.~W. 2008, \aap, 487, 993

\bibitem[{{Kounkel}(2020)}]{Kounkel2020}
{Kounkel}, M. 2020, \apj, 902, 122

\bibitem[{{Kounkel} {et~al.}(2018){Kounkel}, {Covey}, {Su{\'a}rez},
  {Rom{\'a}n-Z{\'u}{\~n}iga}, {Hernandez}, {Stassun}, {Jaehnig}, {Feigelson},
  {Pe{\~n}a Ram{\'\i}rez}, {Roman-Lopes}, {Da Rio}, {Stringfellow}, {Kim},
  {Borissova}, {Fern{\'a}ndez-Trincado}, {Burgasser},
  {Garc{\'\i}a-Hern{\'a}ndez}, {Zamora}, {Pan}, \& {Nitschelm}}]{Kounkel2018}
{Kounkel}, M., {Covey}, K., {Su{\'a}rez}, G., {et~al.} 2018, \aj, 156, 84

\bibitem[{{Kristensen} {et~al.}(2010){Kristensen}, {van Dishoeck}, {van
  Kempen}, {Cuppen}, {Brinch}, {J{\o}rgensen}, \&
  {Hogerheijde}}]{Kristensen2010}
{Kristensen}, L.~E., {van Dishoeck}, E.~F., {van Kempen}, T.~A., {et~al.} 2010,
  \aap, 516, A57

\bibitem[{{Lada} \& {Black}(1976)}]{Lada1976}
{Lada}, C.~J. \& {Black}, J.~H. 1976, \apjl, 203, L75

\bibitem[{{Ladd} {et~al.}(1998){Ladd}, {Fuller}, \& {Deane}}]{Ladd1998}
{Ladd}, E.~F., {Fuller}, G.~A., \& {Deane}, J.~R. 1998, \apj, 495, 871

\bibitem[{{Lang} {et~al.}(2000){Lang}, {Masheder}, {Dame}, \&
  {Thaddeus}}]{Lang2000}
{Lang}, W.~J., {Masheder}, M.~R.~W., {Dame}, T.~M., \& {Thaddeus}, P. 2000,
  \aap, 357, 1001

\bibitem[{{Lee} {et~al.}(2005){Lee}, {Chen}, {Zhang}, \& {Hu}}]{Lee2005}
{Lee}, H.-T., {Chen}, W.~P., {Zhang}, Z.-W., \& {Hu}, J.-Y. 2005, \apj, 624,
  808

\bibitem[{{Maddalena} \& {Morris}(1987)}]{Maddalena1987}
{Maddalena}, R.~J. \& {Morris}, M. 1987, \apj, 323, 179

\bibitem[{{Manigand} {et~al.}(2020){Manigand}, {J{\o}rgensen}, {Calcutt},
  {M{\"u}ller}, {Ligterink}, {Coutens}, {Drozdovskaya}, {van Dishoeck}, \&
  {Wampfler}}]{Manigand2020}
{Manigand}, S., {J{\o}rgensen}, J.~K., {Calcutt}, H., {et~al.} 2020, \aap, 635,
  A48

\bibitem[{{Mart{\'\i}n-Dom{\'e}nech} {et~al.}(2016){Mart{\'\i}n-Dom{\'e}nech},
  {Mu{\~n}oz Caro}, \& {Cruz-D{\'\i}az}}]{Martin-Domenech2016}
{Mart{\'\i}n-Dom{\'e}nech}, R., {Mu{\~n}oz Caro}, G.~M., \& {Cruz-D{\'\i}az},
  G.~A. 2016, \aap, 589, A107

\bibitem[{{Mathieu}(2008)}]{Mathieu2008}
{Mathieu}, R.~D. 2008, {The {\ensuremath{\lambda}} Orionis Star Forming
  Region}, Vol. 4, ASP Monographs ({Reipurth}, B. ed), 757

\bibitem[{{McGuire}(2018)}]{McGuire2018}
{McGuire}, B.~A. 2018, \apjs, 239, 17

\bibitem[{{McMullin} {et~al.}(2007){McMullin}, {Waters}, {Schiebel}, {Young},
  \& {Golap}}]{McMullin2007}
{McMullin}, J.~P., {Waters}, B., {Schiebel}, D., {Young}, W., \& {Golap}, K.
  2007, in Astronomical Society of the Pacific Conference Series, Vol. 376,
  Astronomical Data Analysis Software and Systems XVI, ed. R.~A. {Shaw},
  F.~{Hill}, \& D.~J. {Bell}, 127

\bibitem[{{Morgan} {et~al.}(2008){Morgan}, {Thompson}, {Urquhart}, \&
  {White}}]{Morgan2008}
{Morgan}, L.~K., {Thompson}, M.~A., {Urquhart}, J.~S., \& {White}, G.~J. 2008,
  \aap, 477, 557

\bibitem[{{M{\"u}ller} {et~al.}(2001){M{\"u}ller}, {Thorwirth}, {Roth}, \&
  {Winnewisser}}]{Muller2001}
{M{\"u}ller}, H.~S.~P., {Thorwirth}, S., {Roth}, D.~A., \& {Winnewisser}, G.
  2001, \aap, 370, L49

\bibitem[{{Murdin} \& {Penston}(1977)}]{Murdin1977}
{Murdin}, P. \& {Penston}, M.~V. 1977, \mnras, 181, 657

\bibitem[{{Myers} {et~al.}(1988){Myers}, {Heyer}, {Snell}, \&
  {Goldsmith}}]{Myers1988}
{Myers}, P.~C., {Heyer}, M., {Snell}, R.~L., \& {Goldsmith}, P.~F. 1988, \apj,
  324, 907

\bibitem[{{Myers} {et~al.}(1983){Myers}, {Linke}, \& {Benson}}]{Myers1983}
{Myers}, P.~C., {Linke}, R.~A., \& {Benson}, P.~J. 1983, \apj, 264, 517

\bibitem[{{Neufeld} \& {Dalgarno}(1989)}]{Neufeld1989}
{Neufeld}, D.~A. \& {Dalgarno}, A. 1989, \apj, 340, 869

\bibitem[{{Nisini} {et~al.}(2010){Nisini}, {Benedettini}, {Codella},
  {Giannini}, {Liseau}, {Neufeld}, {Tafalla}, {van Dishoeck}, {Bachiller},
  {Baudry}, {Benz}, {Bergin}, {Bjerkeli}, {Blake}, {Bontemps}, {Braine},
  {Bruderer}, {Caselli}, {Cernicharo}, {Daniel}, {Encrenaz}, {di Giorgio},
  {Dominik}, {Doty}, {Fich}, {Fuente}, {Goicoechea}, {de Graauw}, {Helmich},
  {Herczeg}, {Herpin}, {Hogerheijde}, {Jacq}, {Johnstone}, {J{\o}rgensen},
  {Kaufman}, {Kristensen}, {Larsson}, {Lis}, {Marseille}, {McCoey}, {Melnick},
  {Olberg}, {Parise}, {Pearson}, {Plume}, {Risacher}, {Santiago}, {Saraceno},
  {Shipman}, {van Kempen}, {Visser}, {Viti}, {Wampfler}, {Wyrowski}, {van der
  Tak}, {Y{\i}ld{\i}z}, {Delforge}, {Desbat}, {Hatch}, {P{\'e}ron}, {Schieder},
  {Stern}, {Teyssier}, \& {Whyborn}}]{Nisini2010}
{Nisini}, B., {Benedettini}, M., {Codella}, C., {et~al.} 2010, \aap, 518, L120

\bibitem[{{Noble} {et~al.}(2012){Noble}, {Congiu}, {Dulieu}, \&
  {Fraser}}]{Noble2012}
{Noble}, J.~A., {Congiu}, E., {Dulieu}, F., \& {Fraser}, H.~J. 2012, \mnras,
  421, 768

\bibitem[{{Noble} {et~al.}(2013){Noble}, {Fraser}, {Aikawa}, {Pontoppidan}, \&
  {Sakon}}]{Noble2013}
{Noble}, J.~A., {Fraser}, H.~J., {Aikawa}, Y., {Pontoppidan}, K.~M., \&
  {Sakon}, I. 2013, \apj, 775, 85

\bibitem[{{Noble} {et~al.}(2017){Noble}, {Fraser}, {Pontoppidan}, \&
  {Craigon}}]{Noble2017}
{Noble}, J.~A., {Fraser}, H.~J., {Pontoppidan}, K.~M., \& {Craigon}, A.~M.
  2017, Monthly Notices of the Royal Astronomical Society, 467, 4753

\bibitem[{{\"O}berg(2016)}]{Oberg2016}
{\"O}berg, K.~I. 2016, Chemical Reviews, 116, 9631, pMID: 27099922

\bibitem[{{{\"O}berg} \& {Bergin}(2021)}]{Oberg2020}
{{\"O}berg}, K.~I. \& {Bergin}, E.~A. 2021, \physrep, 893, 1

\bibitem[{{{\"O}berg} {et~al.}(2009){{\"O}berg}, {Bottinelli}, \& {van
  Dishoeck}}]{Oberg2009a}
{{\"O}berg}, K.~I., {Bottinelli}, S., \& {van Dishoeck}, E.~F. 2009, \aap, 494,
  L13

\bibitem[{{{\"O}berg} {et~al.}(2011){{\"O}berg}, {Murray-Clay}, \&
  {Bergin}}]{Oberg2011}
{{\"O}berg}, K.~I., {Murray-Clay}, R., \& {Bergin}, E.~A. 2011, \apjl, 743, L16

\bibitem[{{Ossenkopf} \& {Henning}(1994)}]{Ossenkopf1994}
{Ossenkopf}, V. \& {Henning}, T. 1994, \aap, 291, 943

\bibitem[{{Penteado} {et~al.}(2017){Penteado}, {Walsh}, \&
  {Cuppen}}]{Penteado2017}
{Penteado}, E.~M., {Walsh}, C., \& {Cuppen}, H.~M. 2017, \apj, 844, 71

\bibitem[{{Perotti} {et~al.}(2020){Perotti}, {Rocha}, {J{\o}rgensen},
  {Kristensen}, {Fraser}, \& {Pontoppidan}}]{Perotti2020}
{Perotti}, G., {Rocha}, W.~R.~M., {J{\o}rgensen}, J.~K., {et~al.} 2020, \aap,
  643, A48

\bibitem[{{Pickett} {et~al.}(1998){Pickett}, {Poynter}, {Cohen}, {Delitsky},
  {Pearson}, \& {M{\"u}ller}}]{Pickett1998}
{Pickett}, H.~M., {Poynter}, R.~L., {Cohen}, E.~A., {et~al.} 1998, \jqsrt, 60,
  883

\bibitem[{{Qin} \& {Wu}(2003)}]{Qin2003}
{Qin}, S.-L. \& {Wu}, Y.-F. 2003, \cjaa, 3, 69

\bibitem[{{Rabli} \& {Flower}(2010)}]{Rabli2010}
{Rabli}, D. \& {Flower}, D.~R. 2010, \mnras, 406, 95

\bibitem[{{Reipurth}(2000)}]{Reipurth2000}
{Reipurth}, B. 2000, VizieR Online Data Catalog, V/104

\bibitem[{{Reipurth} \& {Bally}(2001)}]{Reipurth2001}
{Reipurth}, B. \& {Bally}, J. 2001, \araa, 39, 403

\bibitem[{{Reipurth} \& {Friberg}(2021)}]{Reipurth2020}
{Reipurth}, B. \& {Friberg}, P. 2021, \mnras, 501, 5938

\bibitem[{{Sahan} \& {Haffner}(2016)}]{Sahan2016}
{Sahan}, M. \& {Haffner}, L.~M. 2016, \aj, 151, 147

\bibitem[{{Santangelo} {et~al.}(2012){Santangelo}, {Nisini}, {Giannini},
  {Antoniucci}, {Vasta}, {Codella}, {Lorenzani}, {Tafalla}, {Liseau}, {van
  Dishoeck}, \& {Kristensen}}]{Santangelo2012}
{Santangelo}, G., {Nisini}, B., {Giannini}, T., {et~al.} 2012, \aap, 538, A45

\bibitem[{{Sch{\"o}ier} {et~al.}(2005){Sch{\"o}ier}, {van der Tak}, {van
  Dishoeck}, \& {Black}}]{Schoier2005}
{Sch{\"o}ier}, F.~L., {van der Tak}, F.~F.~S., {van Dishoeck}, E.~F., \&
  {Black}, J.~H. 2005, \aap, 432, 369

\bibitem[{{Sharpless}(1959)}]{Sharpless1959}
{Sharpless}, S. 1959, \apjs, 4, 257

\bibitem[{{Skrutskie} {et~al.}(2006){Skrutskie}, {Cutri}, {Stiening},
  {Weinberg}, {Schneider}, {Carpenter}, {Beichman}, {Capps}, {Chester},
  {Elias}, {Huchra}, {Liebert}, {Lonsdale}, {Monet}, {Price}, {Seitzer},
  {Jarrett}, {Kirkpatrick}, {Gizis}, {Howard}, {Evans}, {Fowler}, {Fullmer},
  {Hurt}, {Light}, {Kopan}, {Marsh}, {McCallon}, {Tam}, {Van Dyk}, \&
  {Wheelock}}]{Skrutskie2006}
{Skrutskie}, M.~F., {Cutri}, R.~M., {Stiening}, R., {et~al.} 2006, \aj, 131,
  1163

\bibitem[{Suutarinen(2015)}]{Suutarinen2015}
Suutarinen, A. 2015, PhD thesis, Dept. of Physics, The Open University,
  \url{http://oro.open.ac.uk/61309/}

\bibitem[{{Suutarinen} {et~al.}(2014){Suutarinen}, {Kristensen}, {Mottram},
  {Fraser}, \& {van Dishoeck}}]{Suutarinen2014}
{Suutarinen}, A.~N., {Kristensen}, L.~E., {Mottram}, J.~C., {Fraser}, H.~J., \&
  {van Dishoeck}, E.~F. 2014, \mnras, 440, 1844

\bibitem[{{van Dishoeck} \& {Blake}(1998)}]{vanDishoeck1998}
{van Dishoeck}, E.~F. \& {Blake}, G.~A. 1998, \araa, 36, 317

\bibitem[{{van Gelder} {et~al.}(2020){van Gelder}, {Tabone}, {Tychoniec}, {van
  Dishoeck}, {Beuther}, {Boogert}, {Caratti o Garatti}, {Klaassen}, {Linnartz},
  {M{\"u}ller}, \& {Taquet}}]{vanGelder2020}
{van Gelder}, M.~L., {Tabone}, B., {Tychoniec}, {\L}., {et~al.} 2020, \aap,
  639, A87

\bibitem[{{van 't Hoff} {et~al.}(2017){van 't Hoff}, {Walsh}, {Kama},
  {Facchini}, \& {van Dishoeck}}]{vantHoff2017}
{van 't Hoff}, M.~L.~R., {Walsh}, C., {Kama}, M., {Facchini}, S., \& {van
  Dishoeck}, E.~F. 2017, \aap, 599, A101

\bibitem[{{Vasta} {et~al.}(2012){Vasta}, {Codella}, {Lorenzani}, {Santangelo},
  {Nisini}, {Giannini}, {Tafalla}, {Liseau}, {van Dishoeck}, \&
  {Kristensen}}]{Vasta2012}
{Vasta}, M., {Codella}, C., {Lorenzani}, A., {et~al.} 2012, \aap, 537, A98

\bibitem[{{Vasyunin} \& {Herbst}(2013)}]{Vasyunin2013}
{Vasyunin}, A.~I. \& {Herbst}, E. 2013, \apj, 769, 34

\bibitem[{{Wade}(1957)}]{Wade1957}
{Wade}, C.~M. 1957, \aj, 62, 148

\bibitem[{{Watanabe} \& {Kouchi}(2002)}]{Watanabe2002}
{Watanabe}, N. \& {Kouchi}, A. 2002, \apjl, 571, L173

\bibitem[{{Weingartner} \& {Draine}(2001)}]{Weingartner2001}
{Weingartner}, J.~C. \& {Draine}, B.~T. 2001, \apj, 548, 296

\bibitem[{{Whittet} {et~al.}(2011){Whittet}, {Cook}, {Herbst}, {Chiar}, \&
  {Shenoy}}]{Whittet2011}
{Whittet}, D.~C.~B., {Cook}, A.~M., {Herbst}, E., {Chiar}, J.~E., \& {Shenoy},
  S.~S. 2011, \apj, 742, 28

\bibitem[{{Wilson}(1999)}]{Wilson1999}
{Wilson}, T.~L. 1999, Reports on Progress in Physics, 62, 143

\bibitem[{{Wilson} \& {Matteucci}(1992)}]{Wilson1992}
{Wilson}, T.~L. \& {Matteucci}, F. 1992, \aapr, 4, 1

\bibitem[{{Wolfire} {et~al.}(1989){Wolfire}, {Hollenbach}, \&
  {Tielens}}]{Wolfire1989}
{Wolfire}, M.~G., {Hollenbach}, D., \& {Tielens}, A.~G.~G.~M. 1989, \apj, 344,
  770

\bibitem[{{Yang} {et~al.}(2010){Yang}, {Stancil}, {Balakrishnan}, \&
  {Forrey}}]{Yang2010}
{Yang}, B., {Stancil}, P.~C., {Balakrishnan}, N., \& {Forrey}, R.~C. 2010,
  \apj, 718, 1062

\bibitem[{{Zhang} {et~al.}(1989){Zhang}, {Laureijs}, {Chlewicki}, {Clark}, \&
  {Wesselius}}]{Zhang1989}
{Zhang}, C.~Y., {Laureijs}, R.~J., {Chlewicki}, G., {Clark}, F.~O., \&
  {Wesselius}, P.~R. 1989, \aap, 218, 231

\bibitem[{{Zhang} {et~al.}(2018){Zhang}, {Romano}, {Ivison}, {Papadopoulos}, \&
  {Matteucci}}]{Zhang2018}
{Zhang}, Z.-Y., {Romano}, D., {Ivison}, R.~J., {Papadopoulos}, P.~P., \&
  {Matteucci}, F. 2018, \nat, 558, 260

\bibitem[{{Zucker} {et~al.}(2020){Zucker}, {Speagle}, {Schlafly}, {Green},
  {Finkbeiner}, {Goodman}, \& {Alves}}]{Zucker2020}
{Zucker}, C., {Speagle}, J.~S., {Schlafly}, E.~F., {et~al.} 2020, \aap, 633,
  A51

\bibitem[{{Zucker} {et~al.}(2019){Zucker}, {Speagle}, {Schlafly}, {Green},
  {Finkbeiner}, {Goodman}, \& {Alves}}]{Zucker2019}
{Zucker}, C., {Speagle}, J.~S., {Schlafly}, E.~F., {et~al.} 2019, \apj, 879,
  125

\end{thebibliography}


\appendix

\section{Production of gas-phase maps}
\label{appendixA}

\subsection{Interferometric and single-dish data combination}
\label{data_combination}
The combined interferometric (SMA) and single-dish (APEX/IRAM~30~m) data were obtained using CASA version 5.6.1 by executing the CASA task {\fontfamily{qcr}\selectfont feather}. The {\fontfamily{qcr}\selectfont feathering} algorithm consists of Fourier transforming and scaling the single-dish (lower resolution) data to the interferometric (higher resolution) data. In a second step, the two datasets with different spatial resolution are merged. In this work, APEX short-spacing are used to correct the CH$_3$OH emission detected by SMA observations, whereas the IRAM~30~m short-spacing are folded into the CO isotopologue emissions observed with the SMA.
The combination of the interferometric and single-dish datasets was carried out following the procedure described in Appendix~B.1 of \citet{Perotti2020} and references therein.

Artefacts might arise when combining interferometric and single dish data. To benchmark our results we have been varying the arguments of the feathering algorithm (e.g., {\fontfamily{qcr}\selectfont sdfactor} (scale factor to apply to single dish image), {\fontfamily{qcr}\selectfont effdishdiam} (new effective single dish diameter) and {\fontfamily{qcr}\selectfont lowpassfiltersd} (filtering out the high spatial frequencies of the single dish image) from the default values. The variations of the above parameters resulted in amplifying the signal close to the primary beam edges by approximately 15\%, therefore the default values were chosen.

\subsection{Optical depth of the CO isotopologues}
\label{otically_thick_emission}

If the emission of the $^{13}$CO and C$^{18}$O isotopologues is co-spatial, the line profiles are similar and the gas ratio abundance $^{12}$CO:$^{13}$CO:C$^{18}$O is constant, the $^{13}$CO optical depth can be assessed from the $^{13}$CO and C$^{18}$O integrated intensity ratio $I(\mathrm{^{13}CO})$/$I(\mathrm{C^{18}O})$ using the formalism firstly described by \citet{Myers1983} and \citet{Ladd1998}. The treatment employed here is adapted from \citet{Myers1983} and uses formula \ref{co_opt_formula} given by e.g., \citet{Carlhoff2013} and \citet{Zhang2018}, relating the integrated intensity ratio of the two isotopologues (Table~\ref{table:co_iso_intensities}) and the optical depth $\tau$:

\begin{equation}
    \frac{I(\mathrm{^{13}CO})}{I(\mathrm{C^{18}O})} \approx \frac{1 - \mathrm{exp}(-\tau^{\mathrm{^{13}CO}})}{1 - \mathrm{exp}(-\tau^{\mathrm{^{13}CO}}/f)}
    \label{co_opt_formula}
.\end{equation}

\noindent In the above equation $\tau^{\mathrm{^{13}CO}}$ is the $\mathrm{^{13}CO}$ optical depth, and $f$ is the intrinsic $\mathrm{^{13}CO}/\mathrm{C^{18}O}$ abundance ratio, which is equal to 7$-$10 for the Milky Way \citep{Wilson1992,Barnes2015}. If $f$=8 is assumed, to be consistent with the approach adopted in \citet{Perotti2020}, $\tau^\mathrm{^{13}CO}$ is equal to 1.22 for B35A$-$2, 0.39 for B35$-$3, 0.47 for B35A$-$5, and 0.03 for B35$-$4, indicating $^{13}$CO optically thick emission towards three (B35A$-$2, B35$-$3 and B35A$-$5) out of four sources in the B35A cloud as $I(\mathrm{^{13}CO})$/$I(\mathrm{C^{18}O})$ is below the intrinsic ratio $f$ for these three sources. The same conclusion is obtained if $f$=7 or $f$=10 are assumed. Based on this result, 
the C$^{18}$O emission is used to estimate the column densities of $^{12}$CO under local thermodynamic equilibrium (LTE) conditions.

\begin{figure}
\centering
\includegraphics[trim={0 0 0 0}, clip, width=\hsize]{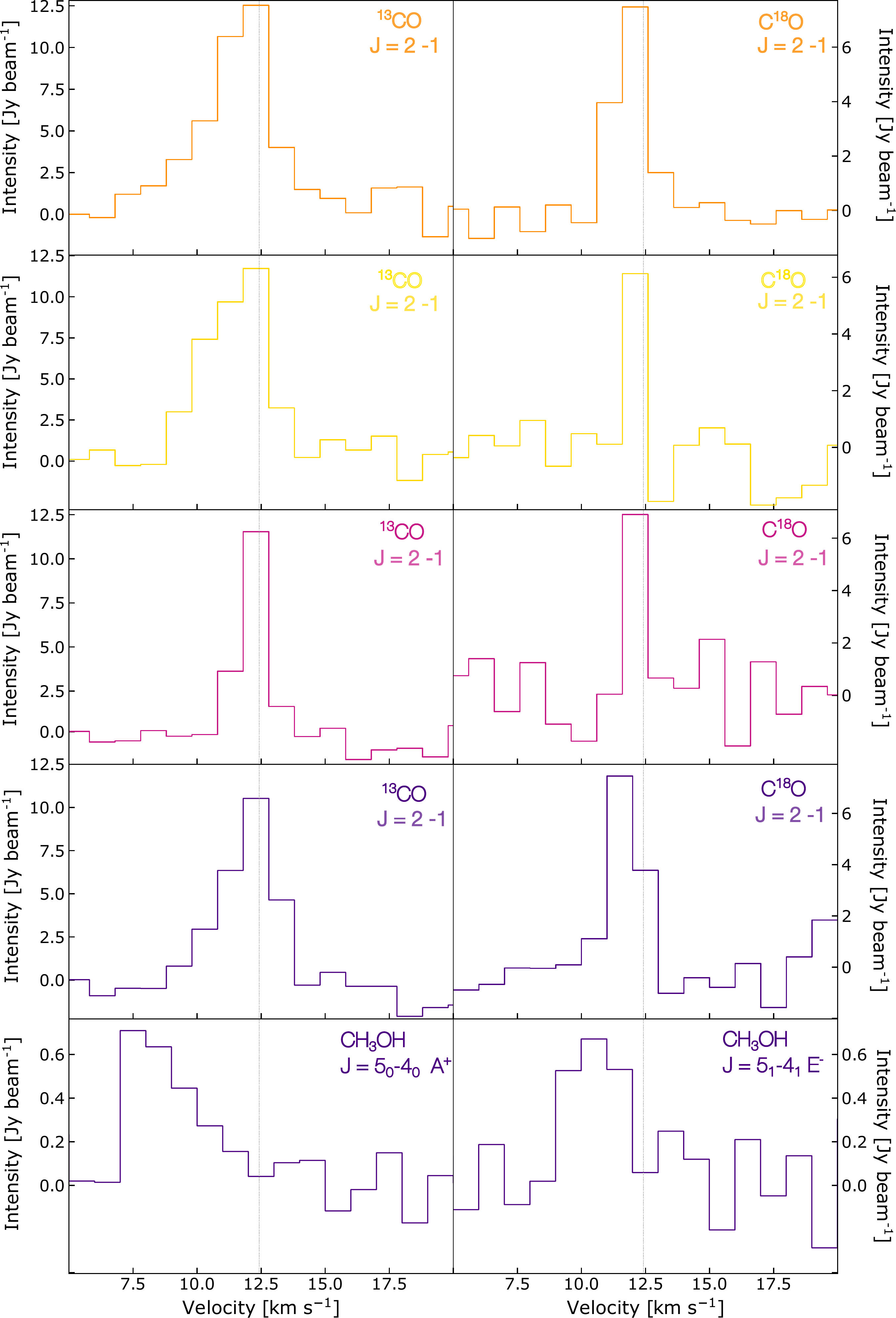}
\caption{$^{13}$CO $J$=2$-$1, C$^{18}$O $J$=2$-$1, CH$_3$OH $J$=5$_0-$4$_0$ A$^{+}$ and CH$_3$OH $J$=5$_1-$4$_1$ E$^{-}$ spectra towards the B35A sources detected in the combined interferometric and single-dish data sets. The spectra are colour-coded according to Figure~\ref{Corr_gas_ice}. The cloud velocity (12.42~km s$^{-1}$) is shown in all spectra with a vertical gray line.}
\label{spectra}
\end{figure}

\begin{table}
\begin{center}
\caption{Integrated $^{13}$CO and C$^{18}$O line intensities in units of Jy~beam$^{-1}$~km~s$^{-1}$ over each source position.}
\label{table:co_iso_intensities}
\renewcommand{\arraystretch}{1.3}
\begin{tabular}{l c c}
\hline \hline
Source & $^{13}$CO ($J = 2-1$)  & C$^{18}$O ($J = 2-1$)\\
\hline
B35A$-$2   & 70.64 $\pm$ 10.59 & 14.18 $\pm$ 2.86  \\
B35A$-$3   & 72.36 $\pm$ 10.85 & 10.66 $\pm$ 2.17  \\
B35A$-$4   & 59.63 $\pm$ 8.95  & 7.55  $\pm$ 1.56  \\
B35A$-$5   & 70.74 $\pm$ 10.61 & 10.75 $\pm$ 2.19  \\
\hline 
\end{tabular}
\end{center}
\begin{center}
\end{center}
\end{table}

\subsection{Channel maps and spectra of individual transitions}
\label{channel maps and spectra}
Channel maps and spectra for the $^{13}$CO $J$=2$-$1, C$^{18}$O $J$=2$-$1 and CH$_3$OH $J$=5$_0-$4$_0$ A$^{+}$ detected in the combined interferometric (SMA) and single-dish (IRAM~30m/APEX) data are displayed in Figures~\ref{spectra}, \ref{ch_maps_13co}, \ref{ch_maps_c18o} and \ref{ch_maps_ch3oh}. The CO isotopologue emission is predominantly concentrated towards B35A$-$2 and B35A$-$3, whereas the CH$_3$OH emission is uniquely localized in the vicinity of B35A$-$5. Blue-shifted components are observed in combination with broad spectral profiles and wings which deviate from the cloud velocity (12.42~km~s$^{-1}$). 

\begin{table*}
\begin{center}
\caption{Integrated CH$_3$OH line intensities in units of Jy~beam$^{-1}$~km~s$^{-1}$ towards B35A$-$5.}
\label{table:ch3oh_intensities}
\renewcommand{\arraystretch}{1.3}
\begin{tabular}{l c c c c c}
\hline \hline
Source & 5$_0~-~4_0$  E$^+$& 5$_1~-~4_1$  E$^-$& 5$_0~-~4_0$  A$^+$& 5$_1~-~4_1$  E$^+$& 5$_2~-~4_2$  E$^-$ \\
\hline
B35A$-$5  &     0.27 $\pm$ 0.10 &  0.77 $\pm$ 0.18  &  1.03 $\pm$ 0.22  &   0.29 $\pm$ 0.11  &  0.20 $\pm$ 0.09    \\
\hline 
\end{tabular}
\end{center}
\end{table*}

\subsection{Derivation of gas-phase column densities} 
\label{gas-phase_cd_formalism}
\begin{figure}
\centering
\includegraphics[trim={0 10 10 0}, clip, width=3.2in]{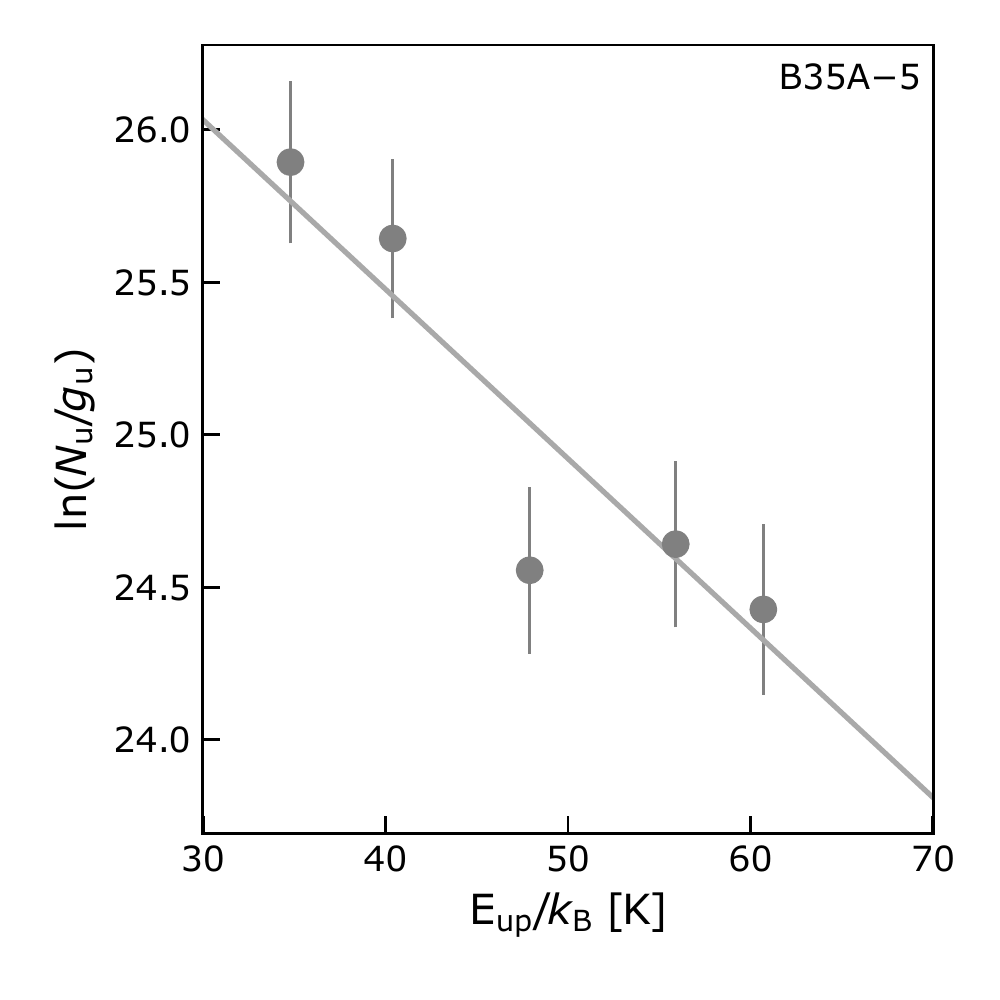}
\caption{Rotational diagram of CH$_3$OH for B35A$-$5. The solid line displays the fixed slope for $T_{\mathrm{rot}}=25$~K. Error bars are for 1$\sigma$ uncertainties. The derived column density is shown in Table~\ref{table:summary_cd}.}
\label{rot_diag}
\end{figure}

\begin{figure*}
\centering
\includegraphics[trim={0 0 0 0}, clip, width=\hsize]{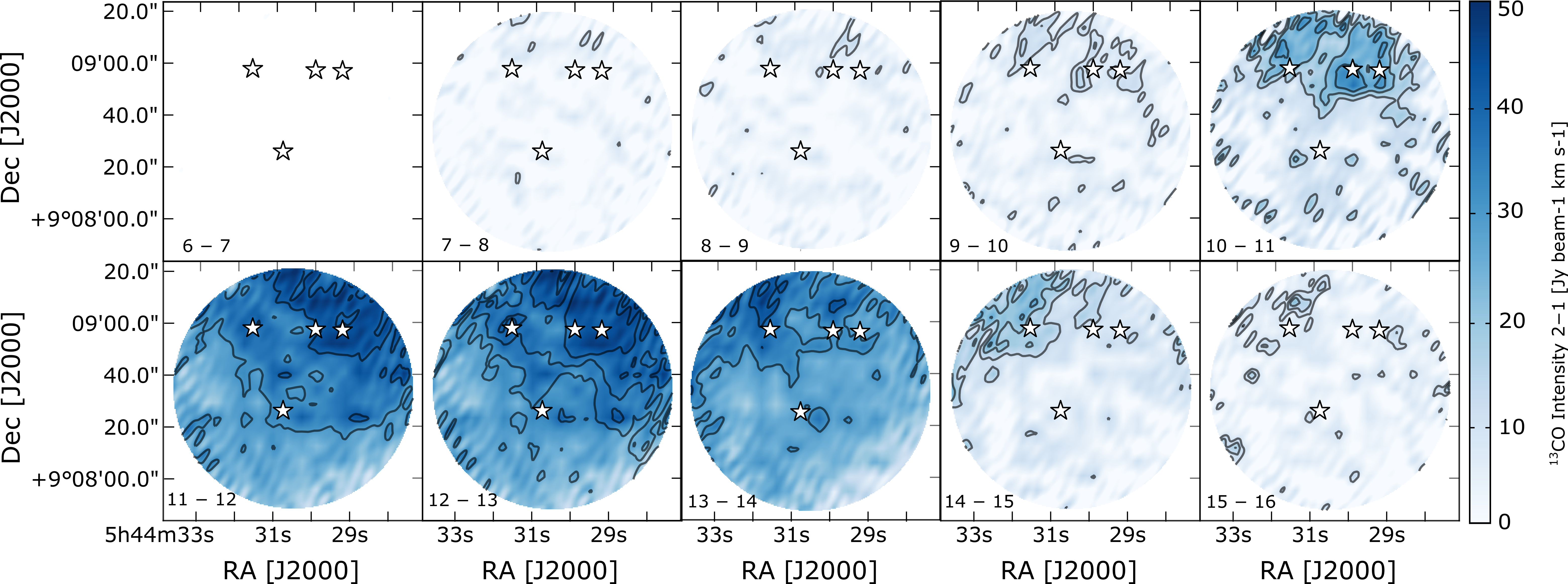}
\caption{Channel maps for $^{13}$CO $J=2-1$ with velocity range 6 to 16~km~s$^{-1}$ in channels of 1~km~s$^{-1}$. Contours starts at 5$\sigma$ and follow a step of 5$\sigma$.}
\label{ch_maps_13co}
\vspace{0.25cm}
\includegraphics[trim={0 0 0 0}, clip, width=\hsize]{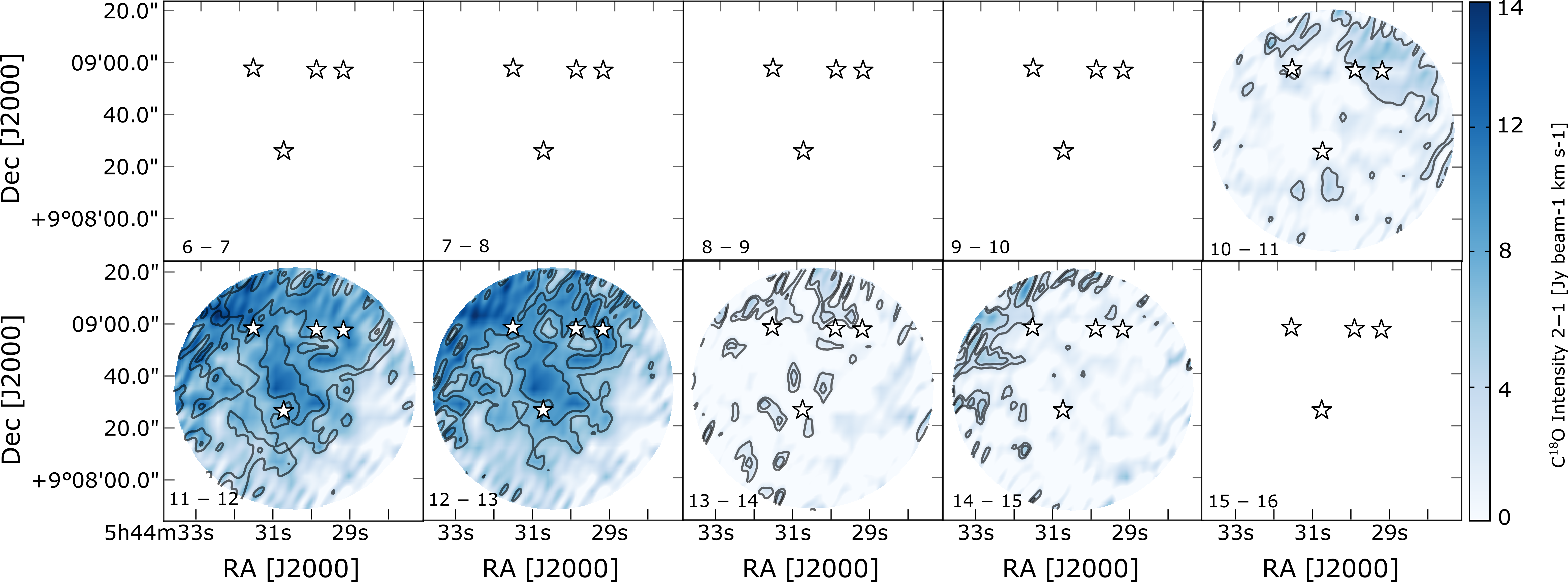}
\caption{Channel maps for C$^{18}$O $J=2-1$ with velocity range 6 to 16~km~s$^{-1}$ in channels of 1~km~s$^{-1}$. Contours starts at 5$\sigma$ and follow a step of 5$\sigma$.}
\label{ch_maps_c18o}
\vspace{0.25cm}
\includegraphics[trim={0 0 0 0}, clip, width=\hsize]{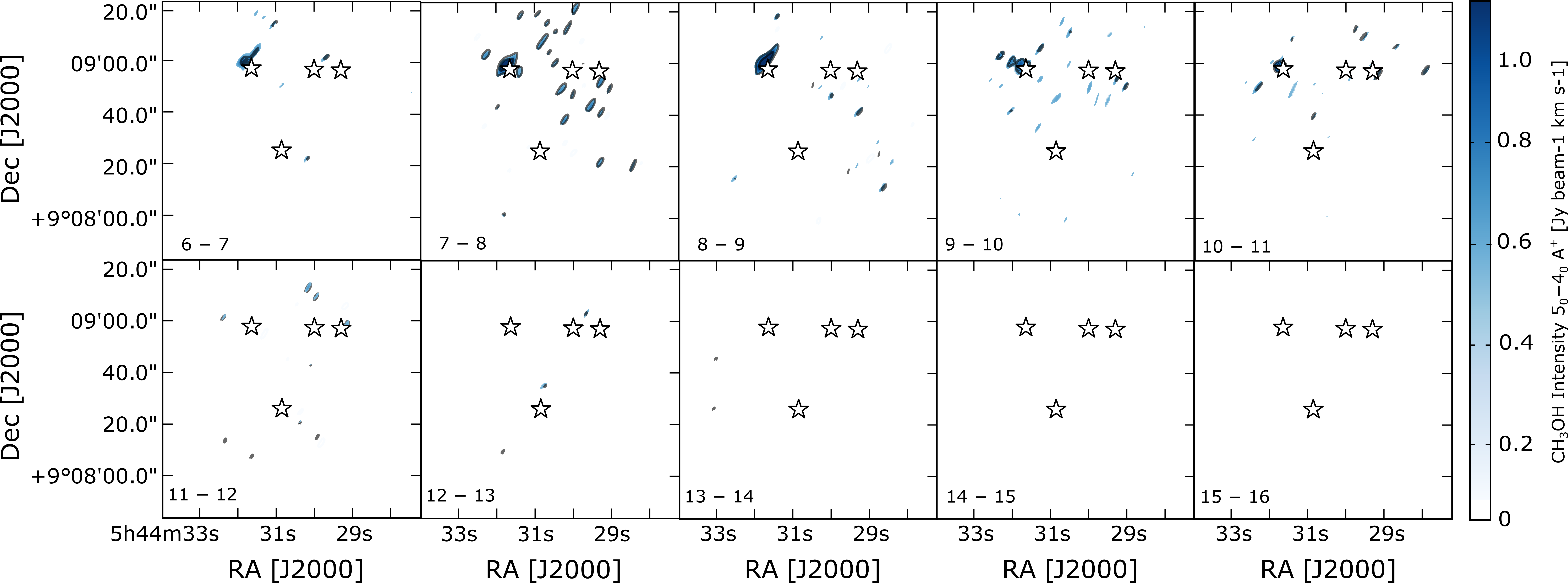}
\caption{Channel maps for CH$_3$OH $J=5_0-4_0$ A$^+$ with velocity range 6 to 16~km~s$^{-1}$ in channels of 1~km~s$^{-1}$. Contours starts at 5$\sigma$ and follow a step of 5$\sigma$.}
\label{ch_maps_ch3oh}
\end{figure*}

The column densities of gas-phase species towards the B35A sources were calculated using the integrated intensities of the combined interferometric (SMA) and single-dish (IRAM~30~m/APEX) data (Tables~ \ref{table:co_iso_intensities} and \ref{table:ch3oh_intensities}). The values inserted in the equations below are listed in Table~\ref{table:spectral_data_pointings} and have been taken from the CDMS \citep{Muller2001}, JPL \citep{Pickett1998} and LAMDA \citep{Schoier2005} spectral databases. The adopted formalism assumes local thermodynamic equilibrium (LTE) conditions and more specifically, optically thin line emission, homogeneous source filling the telescope beam and that level populations can be described by a single excitation temperature. According to the treatment by \citet{Goldsmith1999}, under the aforementioned conditions, the integrated main-beam temperature $\int T_\mathrm{MB}~dv$ and the column density $N_u$ in the upper energy level $u$ are related by:

\begin{equation}
N_u = \frac{8~\pi~k_\mathrm{B}\nu^2}{h~c^3~A_\mathrm{ul}} \int T_\mathrm{MB}~dv
\label{rot_diag1}
\end{equation}

\noindent where $k_\mathrm{B}$ is the Boltzmann's constant, $\nu$ is the transition frequency, $h$ is the Planck's constant, $c$ is the speed of light and $A_\mathrm{ul}$ is the spontaneous Einstein coefficient of the transition. The total column density $N_\mathrm{tot}$ and the rotational temperature $T_\mathrm{rot}$ are related to the column density $N_u$ in the upper energy level $u$ \citep{Goldsmith1999} by: 

\begin{equation}
\frac{N_u}{g_u} = \frac{N_\mathrm{tot}}{Q(T_\mathrm{rot})} e^{-E_u/k_\mathrm{B}T_\mathrm{rot}}
\label{rot_diag2}
\end{equation}

\noindent where $g_u$ is the upper level degeneracy, $Q(T_\mathrm{rot})$ is the rotational partition function, $k_\mathrm{B}$ is the Boltzmann's constant and $E_u$ is the energy of the upper level $u$. In summary, the column densities have been calculated using the formalism of \citet{Goldsmith1999}, but assuming a fixed rotational temperature equal to 25~K \citep{Craigon2015,Reipurth2020}. Figure~\ref{rot_diag} displays the rotational diagram of CH$_3$OH for B35A$-$5. Only CH$_3$OH transitions above 5$\sigma$ were considered. The uncertainties on the gas column densities were estimated based on the rms noise of the spectra and on the $\sim$20\% calibration uncertainty. 

\section{H$_2$ column density from visual extinction}
\label{appendixB}

\begin{figure}
\centering
\includegraphics[trim={0 0 0 0}, clip, width=3.2in]{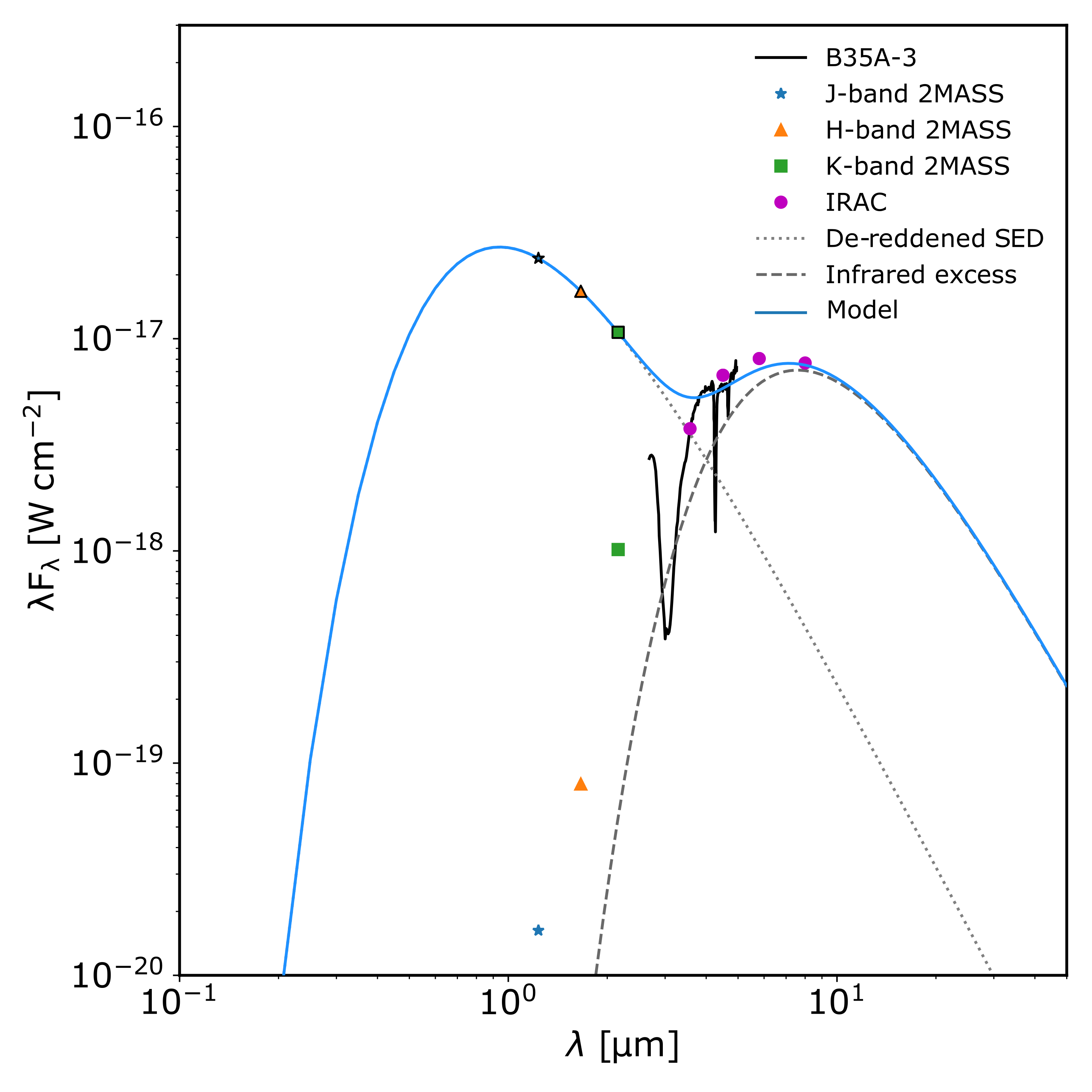}
\caption{The spectral energy distribution (SED) of B35A$-$3, which was modeled to determine its visual extinction. The dotted and dashed lines represent the black body functions used to model the stellar component and the infrared excess, respectively. The solid line is the sum of the two contributions.}
\label{SED_3}
\end{figure}

\begin{table*}
\begin{center}
\caption{Photometry of B35A$-$3.}
\renewcommand{\arraystretch}{1.2}
\label{table:photometry}
\begin{tabular}{lccccccc}
\hline \hline
Object     & 1.235~$\mu$m$^{a}$  & 1.662~$\mu$m$^{a}$  & 2.159~$\mu$m$^{a}$ & 3.6~$\mu$m$^{b}$ & 4.5~$\mu$m$^{b}$ & 5.8~$\mu$m$^{b}$ & 8.0~$\mu$m$^{b}$ \\ 
          &  [mJy]              &       [mJy]         &    [mJy]           &    [mJy]         &     [mJy]        &     [mJy]        &       [mJy]      \\    \hline    
B35A$-$3   &  0.067              &  0.442 $\pm$ 0.0871 &  7.31 $\pm$ 0.229  &  44.59 $\pm$ 2.6 &    101 $\pm$ 6.9 & 156 $\pm$ 8.09   &    205 $\pm$ 9.97  \\    \hline 
\end{tabular}
\end{center}
\footnotesize{\textbf{Notes.} $^{a}$ from 2MASS \citep{Skrutskie2006} $^{b}$ from IRAC \citep{Evans2014}}.
\end{table*}

The production of the H$_2$ column density map is accomplished by using the visual extinction ($A_V$) values for B35A tabulated in the c2d catalog\footnote{\url{https://irsa.ipac.caltech.edu/data/SPITZER/C2D/cores.html}}.
No $A_V$ values are reported for B35A$-$2 and B35A$-$3 in the catalog. As a result, the visual extinction for B35A$-$3 is retrieved by de-reddening its spectral energy distribution (SED) at the J, H, K 2MASS photometric points of Table~\ref{table:photometry} to fit a first blackbody and fit a second blackbody to model the infrared excess at the IRAC photometric points (see Figure~\ref{SED_3}). The extinction in the H-band ($A_H$) is then calculated using the following equation \citep{Chapman2009}: 

\begin{equation}
A_{\lambda} = -2.5~\mathrm{log} \left[ \left(\frac {F_{\lambda}} {B_{\lambda}} \right) \left(\frac {1} {\mathrm{k}} \right)\right]
\label{B3}
\end{equation}

\noindent where $\lambda$ is the wavelength corresponding in this case to the H-band (1.662~$\mu$m), F$_{\lambda}$ is the observed flux at the selected $\lambda$, k is a scaling factor and B$_{\lambda}$ is the blackbody function. The extinction in the H-band is then converted to visual extinction using the equation below: 
\begin{equation}
A_V = 5.55 A_H \mathrm{(mag)} 
\label{B4}
\end{equation}

\noindent where the conversion factor is taken from \citet{Weingartner2001}.
The adopted extinction law takes into account the dust model from \citet{Weingartner2001} for $R_V= 5.5$ designed for the dense interstellar medium and used by the c2d collaboration \citep{Chapman2009, Evans2009}. 
No near-IR photometry data are available for B35A$-$2, thus the $A_V$ for this source is obtained by interpolating all the tabulated visual extinction values for B35A taken from the c2d catalog including the $A_V$ value for B35A$-$3. The obtained visual extinction map of B35A is then converted to a $H_2$ column density map using the relation:

\begin{equation}
N_\mathrm{H_2}= 1.37 \times 10^{21} \mathrm {cm}^{-2} (A_V\mathrm{/mag})
\end{equation}

\noindent established for dense interstellar medium gas \citep{Evans2009}. 

\end{document}